\documentclass[11pt,a4paper]{article}

\usepackage[utf8]{inputenc}  
\usepackage[english]{babel}
\usepackage[T1]{fontenc}
\usepackage{indentfirst}
\usepackage{natbib} 
\usepackage{graphicx}
\usepackage{breqn}
\usepackage{amssymb}
\usepackage{latexsym}
\usepackage[labelfont=bf]{caption}
\usepackage[hidelinks]{hyperref}
\hypersetup{
    colorlinks,
    linkcolor={red!50!black},
    citecolor={blue!50!black},
    urlcolor={blue!80!black}
}
\usepackage{rotating}
\usepackage{collcell}
\usepackage{datatool}
\usepackage{multirow}
\usepackage{array,colortbl}
\usepackage{palatino} % Use the Palatino font by default

\usepackage{float}

\usepackage{geometry}
\usepackage[makeroom]{cancel}
\usepackage{soul}
\usepackage[title]{appendix}

\makeatletter
\def\ps@pprintTitle{%
 \let\@oddhead\@empty
 \let\@evenhead\@empty
 \def\@oddfoot{}%
 \let\@evenfoot\@oddfoot}
\makeatother

\providecommand{\keywords}[1]{\textbf{\textit{Keywords: }} #1}

\usepackage{authblk}

\usepackage{soul}

\newcommand{\virg}[1]{``#1''}

\title{Combining Value-at-Risk and Expected Shortfall forecasts via the Model Confidence Set}
\date{\today}
\author{A. Amendola\thanks{alamendola@unisa.it}}
\author{V. Candila\thanks{vcandila@unisa.it}}
\author{A. Naimoli\thanks{anaimoli@unisa.it}}
\author{G. Storti\thanks{storti@unisa.it}}
\affil{Department of Economics and Statistics, University of Salerno, Italy}

\usepackage{subcaption}
\usepackage[flushleft]{threeparttable}
\usepackage{adjustbox}
\usepackage{arydshln}
\usepackage{bbm}
\usepackage{bm}
\usepackage{amsmath}

\usepackage{soul}

\DeclareSymbolFont{yhlargesymbols}{OMX}{yhex}{m}{n} \DeclareMathAccent{\yhwidehat}{\mathord}{yhlargesymbols}{"62}

\usepackage{booktabs,dcolumn}
\usepackage[table]{xcolor}

\newcolumntype{.}{D{.}{.}{-1}}
\usepackage{floatpag}
\usepackage[normalem]{ulem}

\begin{document}
 
\maketitle

\begin{abstract}
\noindent
To comply with increasingly stringent international standards in risk management and regulation, several approaches have been developed in the literature for forecasting tail-risk measures such as Value-at-Risk (VaR) and Expected Shortfall (ES).
However, the accuracy of these measures can be significantly affected by multiple sources of uncertainty, including model misspecification, data limitations and estimation procedures.
To address these challenges and enhance the predictive performance of individual models, this study introduces novel forecast combination strategies based on the Model Confidence Set (MCS) methodology. Specifically, a strictly consistent joint VaR-ES loss function is employed to identify the best-performing models, which constitute the Set of Superior Models (SSM).
Subsequently, the VaR and ES forecasts of the models included in the SSM are combined using various weighting schemes.
An empirical analysis based on nine stock market indices at the 2.5\% and 1\% risk levels provides evidence that the proposed combined predictors are a robust alternative for forecasting tail-risk measures, successfully passing standard backtests and consistently entering the SSM of the MCS.

\end{abstract}


\section{Introduction}
\label{sec:intro}
\noindent
The ongoing evolution of socio-economic policy scenarios and events that continuously influence the performance of financial markets makes tail risk a crucial concern for risk management and financial regulation.
Value-at-Risk (VaR) and Expected Shortfall (ES) are the main risk measures used for regulatory capital calculation, decision making and risk management in the current Basel III banking regulatory framework. 

From a risk management perspective, understanding the relative forecasting efficiency of alternative models, the temporal stability of their predictive performance and their ability to promptly adapt to changing market conditions are of fundamental importance.
However, these evaluations are inherently complex and often inconclusive, due to the numerous sources of uncertainty that can influence VaR and ES forecasts.

First, several approaches, parametric, non-parametric and semi-parametric, can be used to generate VaR and ES forecasts.
Parametric methods often refer to GARCH models and require the conditional distribution of returns and volatility dynamics to be specified.
On the other hand, semi-parametric models, such as quantile regression \citep{Koenker:Bassett:1978, Engle:Manganelli:2004, taylor2019forecasting, Gerlach:Wang:2020} and Filtered Historical Simulation \citep{barone1999var}, make assumptions on the dynamics of risk rather than on the full conditional distribution of returns.
Finally, the historical simulation \citep[HS,][]{Hendricks:1996} is a popular example of a non-parametric approach.
Despite its simplicity, some theoretical inconsistencies have limited the diffusion of this approach among researchers and analysts \citep[see e.g.][]{holton2013var}.

Regardless of the class of models used, there exist other sources of uncertainty that can significantly affect the accuracy of tail risk forecasts.
Deciding whether to use low-frequency information from daily returns or high-frequency information from intraday data introduces a first source of uncertainty.
Although using high-frequency data allows for improved volatility and tail risk forecasts \citep{andersen2003modeling, hansen2011forecasting, gerlach2020time}, identifying the optimal sampling  frequency of intraday returns could be challenging, due to the effect of jumps and market micro-structure noise on observed intraday prices.
In this context, it is beneficial for tail risk forecasting to consider interactions among different realized volatility measures characterized by different sensitivities to noise and jumps along with combinations of realized measures obtained from multiple sampling frequencies \citep{naimoli2021improving}. 

In addition, while some models can accurately predict both VaR and ES during periods of high volatility, these models may perform poorly during less turbulent periods, making tail risk forecasting highly data dependent.
A possible solution to deal with this additional source of uncertainty is to consider time-varying parameter models that enable volatility dynamics to adapt to changing market conditions \citep{Bollerslev:Patton:Quaedvlieg:2016, gerlach2020time}.

Also, for the class of parametric and semi-parametric models, the choice of the estimation method introduces another source of uncertainty. While this issue has been explored in the literature on the estimation of high-dimensional conditional covariance matrices \citep[see, e.g.,][among others]{fan2008modelling, li2016modeling, engle2019large, pakel2021fitting}, it is still at an early stage of investigation for VaR and ES estimation methods.

Finally, exogenous factors such as sentiment and attention measures can have a significant impact on stock market volatility and tail risk forecasts \citep{Audrino2020, Xu2021, li2022forecasting, NAIMOLI2023}.

From the previous discussion, it is clear that the determination of the ideal forecasting model is subject to data, parameters and model uncertainties. To minimize the impact of uncertainty sources when the optimal model is unknown or may change in the future, a potentially effective solution is to combine forecasts from multiple methods. Regardless of the area of application (economics, finance, physical, environmental sciences, sport, etc.), using forecast combination reduces the risk of relying on a single model and potentially increases forecast accuracy.
In fact, using only one (or a subset of) model(s) can lead to poor performance compared to some combined predictors that, adaptively and dynamically, select only the best-performing models. 

%
% Literature review section
%

Since the pioneering work of \cite{BatesGranger1969}, a substantial number of studies have shown that combining forecasts is a beneficial practice that produces, on average, significantly better predictions than individual models. A comprehensive review of the combination literature can be found in \cite{Wang:Hyndman:Li:Kang:2023}. While most of the forecast combination literature has focused on conditional mean or volatility forecasts, relatively little attention has so far been devoted to the joint combination of VaR and ES.

Focusing on VaR, as no single model or approach dominates forecast comparisons \citep[see][among others]{Komunjer:2013,Nieto:Ruiz:2016}, a number of studies have provided empirical evidence in favor of combining VaR forecasts using different strategies to synthesize information from individual models. \cite{Halbleib:Pohlmeier:2012} developed methods for the combination of VaR forecasts by means of conditional coverage optimization and quantile regression (QR). To select a reliable VaR prediction regardless of the time period, \cite{mcaleer2013gfc} examine twelve novel approaches that combine VaR forecasts obtained using univariate models. These approaches include lower-bound, upper-bound and average methods, in addition to nine strategies based on percentiles ranging from the 10th to the 90th, including the median.
\cite{FUERTES201328} also applied QR to combine VaR forecasts from inter-day and intra-day models and to develop a Wald-type conditional quantile forecast encompassing test.
\cite{Jeron:Taylor:2013} introduced forecasting combination methods that use QR for weight estimation, extending the Conditional Autoregressive Value at Risk (CAViaR) model of \cite{Engle:Manganelli:2004} by adding implied volatility as an additional predictor.
Finally, \cite{BAYER2018} proposed combining VaR forecasts with penalized QR, considering regularization by ridge, lasso and elastic net penalties.

Nevertheless, the literature on combining VaR and ES is still in its early stages. To the best of our knowledge, only \cite{Taylor:2020} and \cite{Storti:Wang:2023} have presented approaches for jointly combining VaR and ES forecasts. 
\cite{Taylor:2020} proposed two forecast strategies for combining joint (VaR, ES) models: the Minimum Score combining (MS-Comb) and the Relative Score combining (RS-Comb). 
The RS-Comb predictor is a weighted average of the candidate risk forecasts where the weight assigned to each model is a non-linear function of the associated forecast loss. Differently, in the MS-Comb approach, the combination weights are estimated by minimizing
a strictly consistent joint VaR and ES loss function ($FZLoss$) belonging to the class derived by \cite{Fissler:Ziegel:2016}.  \cite{Storti:Wang:2023} proposed a novel methodology for predicting VaR and ES using forecast combination and weighted quantile techniques.

Forecast combination methods have proven to be effective at risk forecasting in many settings. Nevertheless, they suffer from some potential limitations that can affect their performances in some specific settings. First, the inclusion in the universe of poorly performing predictors can have a negative impact on the accuracy of  the combined forecasts. Second, when combining a large number of predictors, the implementation of optimization-based methods, such as the MS-Comb, could be problematic due to the large number of estimated parameters (twice the number of models)  and the well known sensitivity of the $FZLoss$ to the choice of initial conditions \citep{taylor2019forecasting}. Finally, some combination procedures such as the one proposed by \cite{Storti:Wang:2023} 
can be computationally intensive when a high number of competing forecasts has to be combined.  In this paper, we propose a novel forecast combination procedure based on the Model Confidence Set \citep[MCS,][]{Hansen+Lunde+Nason:2011} that aims at mitigating the impact of the multiple sources of uncertainty affecting the generation of tail-risk forecasts and overcoming the above discussed limitations and technical difficulties. 
The MCS procedure is also used by \cite{Samuels:Sekkel:2017} as a data-driven trimming device, combining forecasts exclusively within the resulting Set of Superior Models (SSM) in a U.S. macroeconomic application. By contrast, we integrate the MCS within a dedicated tail-risk combination framework, explicitly designed for joint VaR and ES prediction. In particular, in order to determine the best performing models that enter the SSM at each time point, we resort to the use of the $FZLoss$.  We refer to this step as the \emph{training} MCS \citep[as in][]{Amendola2020}.
We adopt two configurations of $FZLoss$. The \textit{unweighted} $FZLoss$, which assigns equal weight to all observations and the \textit{weighted} $FZLoss$, which places more emphasis on recent observations.
The idea of giving more weight to recent observations has been previously investigated by \cite{taylor2008exponentially}, among others. In the proposed combination strategies, the weighted $FZLoss$ is based on an exponential smoothing approach using an exogenously determined decay parameter. 
The VaR and ES forecasts from the models included in the SSM are then used to generate combined risk predictors according to three different weighting schemes: simple average, the RS-Comb and MS-Comb methods. Overall, six different MCS-based combined predictors are proposed.

The performance of our proposed MCS-based combined predictors is compared to each individual model
and four alternative combined predictors computed over the whole universe of VaR and ES forecasts: the mean and median forecasts, in the spirit of \cite{mcaleer2013gfc} and the global RS-Comb and MS-Comb predictors.

For predictive analysis assessment, we implement backtesting using the Unconditional Coverage (UC) test according to \cite{kupiec1995techniques}, the Conditional Coverage (CC) test following \cite{Christoffersen:1998} and the Dynamic Quantile (DQ) test developed by \cite{Engle:Manganelli:2004}, in addition to the Regression-Based Expected Shortfall Backtesting (BD) method presented by \cite{Bayer:Dimitriadis:2020}. Furthermore, we evaluate the forecasting accuracy of the competing predictors by means of the $FZLoss$. The MCS is then used to identify the SSM over the whole forecasting period. We refer to this step as the \emph{evaluation} MCS.

The empirical application extensively evaluates the performance of the combined MCS-based predictors on nine stock market indices, namely S\&P 500, Shanghai Composite, Euro Stoxx 50, NASDAQ, Nikkei, Hang Seng, MXX, BSESN, and Bovespa. For these indices, the VaR and ES at the 2.5\% and 1\% levels are generated using 32 parametric, non-parametric and semi-parametric approaches. The forecasts are then combined according to the proposed predictors.

The results show that, although no single model consistently passes all backtesting procedures or is always included in the \emph{evaluation} MCS across all indices, the proposed combinations generally demonstrate robust performance. Several proposed MCS-based combinations are consistently included in the MCS across most indices, particularly at the 2.5\% risk level, and frequently pass the backtesting procedures at both coverage levels.

%\textcolor{red}{QQQ: Che ne dite? 
%The contribution of this paper is twofold. First, we extend the MCS-based approach to the VaR-ES forecasting framework by proposing six combined predictors that integrate model selection and combination, mitigating multiple sources of uncertainty in tail-risk forecasts. Second, we show that these combined predictors provide statistically significant improvements over standard benchmarks across different markets and coverage levels.
%}

%This paper extends the MCS-based approach to the VaR–ES forecasting framework by proposing six combined predictors that integrate model selection and combination, with the aim of mitigating multiple sources of uncertainty in tail-risk forecasts. The empirical findings indicate that the proposed combined predictors yield statistically significant improvements over standard benchmarks across different markets and coverage levels.

The remaining sections are organized as follows. Section \ref{sec:environment} describes the statistical framework and Section \ref{sec:comb_predict} introduces the proposed combined predictors. Section \ref{sec:emp_analysis} presents an empirical application to six stock market indices and Section \ref{sec:conclusions} concludes.

\section{Description of the environment \label{sec:environment}}

Let $r_{i}$ denote the daily log-returns of a given asset on day $i$, computed as the first difference of log--prices. Moreover, let the information set available at time $i$ be $\mathcal{F}_{i}$. We assume that $r_{i}$ follows: 
\begin{equation}
r_{i} =\sqrt{h_{i}}  \eta_{i}, \quad \text{with} \quad  \eta_{i}\overset{i.i.d}{\sim} \left(0, 1\right),
	\label{eq:cond_dist}
\end{equation}
where $h_{i}=var(r_{i}|\mathcal{F}_{i-1} )$ is the conditional variance of $r_i$ and  $\eta_{i}$ has a cumulative distribution function denoted by $F(\cdot)$ that is assumed to be strictly increasing and continuous on the real line $\Re$.

The (one-step-ahead) VaR for day $i$  at $\tau$ level for $r_{i}$, denoted as $VaR_{i}(\tau)$, is then defined as the conditional $\tau$  quantile of $r_{i}$ that is:
$$
Pr(r_{i} \leq VaR_{i}(\tau)|\mathcal{F}_{i-1})= \tau,
$$
therefore: 
\begin{equation}
VaR_{i}(\tau)\equiv Q_{r_{i}}\left(\tau|\mathcal{F}_{i-1}\right)= \sqrt{h_{i}} F^{-1}(\tau),
\label{eq:VaR}
\end{equation}
%where ${F^{-1}(\tau)}= \inf \left\lbrace \eta_{i}: F(\eta_{i}) \geq \tau \right\rbrace$. 
remarking that the notations $VaR_{i}(\tau)$ and $Q_{r_{i}}\left(\tau|\mathcal{F}_{i-1}\right)$ can be used interchangeably. For a given $\tau$, according to the parametric approach, $VaR_{i}(\tau)$ can be obtained first estimating  $h_{i}$ via a dynamic model for the conditional variance of returns \citep[following][for instance]{Engle:1982, Bollerslev:1986} and then retrieving the constant $F^{-1}(\tau)$ parametrically. According to the semi-parametric approach, instead, $Q_{r_{i}}\left(\tau|\mathcal{F}_{i-1}\right)$ can be directly obtained via a quantile regression approach. Finally, in the case of the HS (belonging to the class of non-parametric approaches), $VaR_{i}(\tau)$ is obtained as the empirical quantile of the returns over a rolling window of fixed length. The other risk measure of interest, the ES, is defined as the conditional expectation of the returns when these violate the VaR condition (meaning that the returns are smaller than the  $VaR_{i}(\tau)$). Formally:

\begin{equation}
ES_{i}(\tau)=\mathbb{E}\left[r_{i}|r_{i}\leq VaR_{i}(\tau)\right].
\label{eq:es}
\end{equation}

The availability of strictly consistent loss functions, or \virg{scoring rules}, is essential for the semi-parametric modelling of VaR and ES, and for evaluating the accuracy of their forecasts. Unlike VaR, which is \emph{elicitable} with respect to the quantile loss, there are no strictly consistent scoring functions for ES. However, \cite{Fissler:Ziegel:2016} propose a class of joint loss functions, which are strictly consistent for the pair (VaR, ES), meaning that their expected value is uniquely minimized by the true VaR and ES series.

They show that strictly consistent scoring functions for the joint assessment of VaR and ES have the following general form:
\begin{align}
S_i(r_i, VaR_i(\tau), ES_i(\tau)) 
&= (I_i - \tau) G_1(VaR_i(\tau)) - I_i G_1(r_i) + G_2(ES_i(\tau)) \nonumber \\
& \left( ES_i(\tau) - VaR_i(\tau) + \frac{I_i}{\tau} (VaR_i(\tau) - r_i) \right)- H(ES_i(\tau)) + a(r_i),
\label{eq:joint_loss}
\end{align}
where $I_i=\mathbbm{1}_{(r_{i} \leq VaR_{i}(\tau))}$ is the indicator function
taking value 1 if event occurs and 0 otherwise. 
The function $G_1(\cdot)$ is increasing and $G_2(\cdot)$ is strictly increasing and strictly convex. 
The function $H(\cdot)$ satisfies $H'(x) = G_2(x)$ with $\lim_{x \to -\infty} G_2(x) = 0$, 
while $a(\cdot)$ denotes a real-valued integrable function.
\cite{taylor2019forecasting} points out that assuming $r_i$ to have zero mean and setting $G_1(x)=0$, $G_2(x)=-1/x$, $H(x)=-log(-x)$ and $a=1-log(1-\tau)$, returns a strictly consistent scoring function which is equal to the negative of the log-likelihood function of an Asymmetric Laplace (AL) density, namely:
\begin{equation}
S_i(r_i, VaR_i(\tau), ES_i(\tau)) = 
-\log\left(\frac{\tau-1}{ES_i(\tau)}\right) 
- \frac{(r_i - VaR_i(\tau)) (\tau - I_i)}{\tau \, ES_i(\tau)}.
 \label{eq:AL_loss}
\end{equation}
\cite{Patton:Ziegel:Chen:2019} show that, under the assumption that both VaR and ES are strictly negative, selecting 
$G_1(x)=0$, $G_2(x)=-1/x$ and $a(x)=0$ in Eq. \eqref{eq:joint_loss} generates a loss function, denoted $FZ0$, that is homogeneous of degree zero. This property has been shown, particularly in volatility forecasting applications, to increase the power of \cite{Diebold1995} tests \citep[see also][]{patton2009evaluating}. Accordingly, the $FZ0$ is given by:
\begin{equation}
FZ0_{i}  = \frac{1}{\tau ES_{i}(\tau)} \mathbbm{1}_{(r_{i} \leq VaR_{i}(\tau))}(r_{i} - VaR_{i}(\tau)) + \frac{VaR_{i}(\tau)}{ES_{i}(\tau)} + \log(-ES_{i}(\tau))-1. \label{eq:fzloss}
\end{equation}
The $FZ0$ loss function is employed to identify the best-performing models, which form the SSM during the phase of the \emph{training} MCS procedure.

\section{MCS-based combined predictors \label{sec:comb_predict}}

To facilitate the exposition of the proposed combined predictors, we first introduce the relevant notation.
Let $m \in {1, \dots, M}$ denote a generic model belonging to the $M$-dimensional set of competing models available for forecasting VaR and ES, encompassing parametric, semi-parametric and non-parametric specifications. 

We define as \emph{training} MCS the procedure used to identify the set of \virg{best} models over a rolling window of past forecasts, corresponding to the so-called training period. More specifically, let $\yhwidehat{\mathcal{M}}_{i,T_{in};1-\alpha;M}$ be the SSM at time $i$ among the $M$-dimensional model universe, over a training period based on the last $T_{in}$ observations, with a significance level of $\alpha$. 

The \emph{training} MCS is performed adopting two different loss functions: the \emph{unweighted} version of the $FZ0$ in Eq. \eqref{eq:fzloss}, where all the observations $i$ have the same weight and a \emph{weighted} version of the same loss ($WFZ0_{i}$), where today's observation is computed as a weighted average of past observations, with the aim of putting more emphasis on recent days. The \emph{weighted} version is parametrized as a simple exponential smoother:
\begin{equation}
WFZ0_{i}= \lambda FZ0_{i} + (1-\lambda) WFZ0_{i-1},
\label{eq:wfzloss}
\end{equation}
where $0<\lambda<1$ is a smoothing parameter determining the speed of decay. Such a decay parameter $\lambda$ should be regarded as a hyperparameter of the weighted loss function used during the training step of the procedure, as it governs the exponential rate at which the weights decay. The larger the value of $\lambda$, the greater the weight assigned to the most recent observation. Conversely, for small values of $\lambda$ (e.g., 0.01 or 0.06), the weight is spread over a longer time span. For instance, when $\lambda = 0.06$, as in the empirical application, the sum of the weights of the last 100 observations amounts to 99.70\% of the total weight, implying that the weighted loss depends almost exclusively on these most recent observations.
Placing greater emphasis on recent observations through the weighted loss (WL) in Eq. \eqref{eq:wfzloss} and small values of $\lambda$ allows the selection of the best-performing models in more recent time periods. 
%\textcolor{red}{It is worth noting that estimating the value of $\lambda$ by minimizing a statistical criterion such as the mean squared prediction error would lead to an estimated value that merely summarizes the persistence of the autocorrelation of the loss series, which is not consistent with the role assigned to $\lambda$ in our modelling strategy.}

After selecting the SSM using the \textit{training} MCS with either the \textit{unweighted} or \textit{weighted} $FZ0$ loss, the next step is to combine the VaR and ES forecasts from the models within the SSM.

In this work, six different combined predictors are proposed: MCS-Comb, WL-MCS-Comb, MCS-RS-Comb, WL-MCS-RS-Comb, MCS-MS-Comb and WL-MCS-MS-Comb. The first two predictors (MCS-Comb and WL-MCS-Comb) are based on the simple average of the risk forecasts generated by the models selected at each time point, while the remaining predictors (MCS-RS-Comb, WL-MCS-RS-Comb, MCS-MS-Comb and WL-MCS-MS-Comb) are based on a weighted average of these forecasts. More specifically:

\begin{itemize}
\item \textbf{MCS-Comb}:  
This predictor combines the $VaR$ and $ES$ forecasts of the models belonging to $\yhwidehat{\mathcal{M}}_{i,T_{in};1-\alpha;M}$, identified using the \textbf{\emph{unweighted}} $FZLoss$ in \eqref{eq:fzloss}. The combination is performed by computing an equally weighted average of the risk forecasts generated by the models in the SSM, so that each model contributes uniformly to the final forecast.

\item \textbf{WL-MCS-Comb}:  
This predictor follows the same approach as MCS-Comb, but employs a different model selection criterion. The $\yhwidehat{\mathcal{M}}_{i,T_{in};1-\alpha;M}$ is constructed using the \textbf{\emph{weighted}} $FZLoss$ in \eqref{eq:wfzloss}, which emphasizes recent observations in model evaluation. Again, the combination across models is based on equal weights, so that all models in the selected set contribute equally to the combined forecast.

\item \textbf{MCS-RS-Comb}:  
This predictor extends the MCS-Comb approach by applying the RS-Comb to the models in the $\yhwidehat{\mathcal{M}}_{i,T_{in};1-\alpha;M}$, obtained using the \textbf{\emph{unweighted}} $FZLoss$.

\item \textbf{WL-MCS-RS-Comb}:  
As for the WL-MCS-Comb, this predictor is based on the \textbf{\emph{weighted}} $FZLoss$, ensuring that only models with strong recent performance are selected. However, contrary to WL-MCS-Comb and similarly to MCS-RS-Comb, the WL-MCS-RS-Comb applies the RS-Comb to the models in the $\yhwidehat{\mathcal{M}}_{i,T_{in};1-\alpha;M}$.

\item \textbf{MCS-MS-Comb}:  
This predictor applies the MS-Comb to the models in the $\yhwidehat{\mathcal{M}}_{i,T_{in};1-\alpha;M}$, computed using the \textbf{\emph{unweighted}} $FZLoss$.

\item \textbf{WL-MCS-MS-Comb}:  
This predictor applies MS-Comb to the models in the $\yhwidehat{\mathcal{M}}_{i,T_{in};1-\alpha;M}$, computed using the \textbf{\emph{weighted}} $FZLoss$.

\end{itemize}

To provide a comprehensive overview, we present below the RS-Comb and MS-Comb approaches proposed by \cite{Taylor:2020}. Let $\yhwidehat{VaR}_{i}^{c}(\tau)$ and $\yhwidehat{ES}_{i}^{c}(\tau)$ be the combined predictors of the individual VaR and ES measures at time $i$, denoted by $\yhwidehat{VaR}_{i}^{m}(\tau)$ and $\yhwidehat{ES}_{i}^{m}(\tau)$ for model $m \in \{1,\dots,M\}$.

The RS-Comb assigns a single set of weights to both VaR and ES measures, denoted by $w_m$ for model $m$, which are proportional to past forecast accuracy through:

\begin{equation}
w_m =
\frac{
\exp\!\left(-\psi \sum_{i=1}^{I-1} S\left(\yhwidehat{VaR}_{i}^{m}(\tau), \yhwidehat{ES}_{i}^{m}(\tau), r_i\right)\right)
}{
\sum_{m=1}^{M} \exp\!\left(-\psi \sum_{i=1}^{I-1} S\left(\yhwidehat{VaR}_{i}^{m}(\tau), \yhwidehat{ES}_{i}^{m}(\tau), r_i\right)\right)
},
\end{equation}
where $I$ denotes the number of observations and $S\left(\cdot\right)$ is the chosen joint scoring function, calculated in each period $i$ for each model $m$ and then summed over all $I-1$ in-sample observations. The parameter $\psi > 0$ is a tuning parameter, with values close to zero reducing the method to the simple average.  
The combined forecasts are then given by:
\begin{align}
\yhwidehat{VaR}_{i}^{c}(\tau) & = \sum_{m=1}^{M} w_m \, \yhwidehat{VaR}_{i}^{m}(\tau), \label{eq:var_rs_comb}\\
\yhwidehat{ES}_{i}^{c}(\tau) & = \sum_{m=1}^{M} w_m \,\yhwidehat{ES}_{i}^{m}(\tau) \label{eq:es_rs_comb}.
\end{align}

The parameter $\psi$ can be estimated by optimizing a strictly consistent scoring function. Contrary to the RS-Comb, the MS-Comb method has two distinct sets of weights for the VaR and ES dynamics. In particular, the MS-Comb approach defines the combined VaR and ES forecasts as
\begin{align}
\yhwidehat{VaR}_{i}^{c}(\tau) & = \sum_{m=1}^{M} w^Q_m \,\yhwidehat{VaR}_{i}^{m}(\tau), \\
\yhwidehat{ES}_{i}^{c}(\tau) & = \yhwidehat{VaR}_{i}^{c}(\tau)  + \sum_{m=1}^{M} w^S_m \left( \yhwidehat{ES}_{i}^{m}(\tau) - \yhwidehat{VaR}_{i}^{m}(\tau) \right),
\end{align}
where $w^Q_m$ and $w^S_m$ are convex weights ($w^Q_m, w^S_m \ge 0$, $\forall m$ and $\sum_{m=1}^{M} w^Q_m = \sum_{m=1}^{M} w^S_m
 = 1$) estimated by minimizing a strictly consistent joint scoring function for VaR and ES. In order to account for the sensitivity of initial values in the estimation of the unknown coefficients involved in the RS-Comb and MS-Comb predictors, we have adopted a multi-start initialization procedure in the spirit of \cite{Engle:Manganelli:2004} for the estimation of CAViaR models.

%\textcolor{red}{A standard combination technique, namely the simple average of all the VaR and ES forecasts, is a special case of Eqs. (\ref{eq:var_rs_comb}) and (\ref{eq:es_rs_comb}), obtained by setting $w_m = 1/M$ to assign equal weights (EW) to all models. This combination is hereafter referred to as EW-Comb. Another popular combination technique is the median of all the VaR and ES forecasts, hereafter referred to as Median-Comb.}
%
%\textcolor{red}{While RS-Comb, MS-Comb, EW-Comb, and Median-Comb combine forecasts from all models, the six proposed methods combine VaR and ES forecasts based exclusively on the best-performing models selected by the training MCS procedure.
%} 

A summary of the main features of the six proposed combined predictors is provided in Table \ref{tab:MCS_comb_summary}.

\begin{table}[!h]
\centering
\caption{Comparison of the six combined predictors. \label{tab:MCS_comb_summary}}
\resizebox{\textwidth}{!}{%
\Huge
\setlength{\extrarowheight}{6pt}
\begin{tabular}{cccc}
\toprule
\textbf{Predictor} & \textbf{SSM selection}                                        & \textbf{Combination scheme}                                                                                & \textbf{Interpretation}                                                                                                                             \\ 
\midrule
MCS-Comb           & \begin{tabular}[c]{@{}c@{}}Unweighted\\ $FZLoss$\end{tabular} & Equal weights (EW)                                                                                         & \begin{tabular}[c]{@{}c@{}}All models in the SSM contribute equally \end{tabular}                                 \\ 
           \addlinespace
  \hdashline
   \addlinespace

WL-MCS-Comb        & \begin{tabular}[c]{@{}c@{}}Weighted\\ $FZLoss$\end{tabular}   & Equal weights (EW)                                                                                         & \begin{tabular}[c]{@{}c@{}}Emphasizes recent performance in SSM selection,\\ but still assigns equal weights across selected models\end{tabular}    \\ 

           \addlinespace
  \hdashline
   \addlinespace 

MCS-RS-Comb          & \begin{tabular}[c]{@{}c@{}}Unweighted\\ $FZLoss$\end{tabular} & \begin{tabular}[c]{@{}c@{}} Relative Score Combining (RS-Comb) \end{tabular}          & \begin{tabular}[c]{@{}c@{}}The RS-Comb is applied to all models in the SSM\end{tabular} \\ 

           \addlinespace
  \hdashline
   \addlinespace 
   
WL-MCS-RS-Comb    & \begin{tabular}[c]{@{}c@{}}Weighted\\ $FZLoss$\end{tabular}   & \begin{tabular}[c]{@{}c@{}}Relative Score Combining (RS-Comb)\end{tabular} & \begin{tabular}[c]{@{}c@{}}The RS-Comb is applied to all models in the SSM,\\ with more emphasis on recent performance\end{tabular}                 \\ 

           \addlinespace
  \hdashline
   \addlinespace 

MCS-MS-Comb          & \begin{tabular}[c]{@{}c@{}}Unweighted\\ $FZLoss$\end{tabular} & \begin{tabular}[c]{@{}c@{}} Minimum Score Combining (MS-Comb) \end{tabular}          & \begin{tabular}[c]{@{}c@{}}The MS-Comb is applied to all models in the SSM\end{tabular} \\ 

     \addlinespace
  \hdashline
   \addlinespace 
   
WL-MCS-MS-Comb    & \begin{tabular}[c]{@{}c@{}}Weighted\\ $FZLoss$\end{tabular}   & \begin{tabular}[c]{@{}c@{}}Minimum Score Combining (MS-Comb)\end{tabular} & \begin{tabular}[c]{@{}c@{}}The MS-Comb is applied to all models in the SSM,\\ with more emphasis on recent performance\end{tabular}                 \\ 

\bottomrule

\end{tabular}%
}
\end{table}

Next, we present the algorithm implemented to generate and evaluate the proposed predictors over the out-of-sample period.
Let $j \in \{0,1,2,\dots,nstep-1\}$ be an index denoting the re-estimation step, 
with $nstep$ being both the total number of model re-estimations and the length of the out-of-sample period. 
At each step $j$, the position of the rolling estimation window shifts forward by one observation, 
and new one-step-ahead forecasts are generated accordingly. 
In what follows, the algorithm used to obtain the combined predictors is described:

\begin{enumerate}

\item \textbf{Estimate} all $M$ candidate models on the rolling window with observations from $i=1+j$ to $i=T_{in}+j$.  Conditional on the estimated parameters, generate the \emph{one-step-ahead} VaR and ES forecasts for day $i=T_{in}+j+1$ for all models.
\item \textbf{Compute} the \emph{training} MCS using the (weighted and unweighted) $FZLoss$ over a window consisting of $T_{in}$ observations, defined as follows:
\begin{itemize}
    \item step $j=0$: VaR and ES estimates for the period $i=1,\dots,T_{in}$;
    \item steps $j\ge 1$: VaR and ES estimates for the period $i=1+j,\dots,T_{in}$ \emph{plus} the VaR and ES out-of-sample forecasts for 
    $i=T_{in}+1,\dots,T_{in}+j$ obtained at all previous steps up to $j-1$.
\end{itemize}
\item \textbf{Obtain} the proposed predictors by combining the VaR and ES forecasts of the models in 
      $\yhwidehat{\mathcal{M}}_{i,T_{in};\,1-\alpha;\,M}$ through {MCS-Comb}, {WL-MCS-Comb} and the combinations based on {RS-Comb} and {MS-Comb} (that is, MCS-RS-Comb, WL-MCS-RS-Comb, MCS-MS-Comb and WL-MCS-MS-Comb).
\item \textbf{Iterate} steps 1–3 for $j \in \{0,1,2,\dots,nstep-1\}$.
\end{enumerate}

\begin{figure}[htbp]
\centering
\caption{Summary diagram of the estimation and \emph{training}  MCS procedures \label{fig:summary_diagram}}
		\vspace{-3cm}
		\makebox[\textwidth][c]{\includegraphics[width=1.25\textwidth]{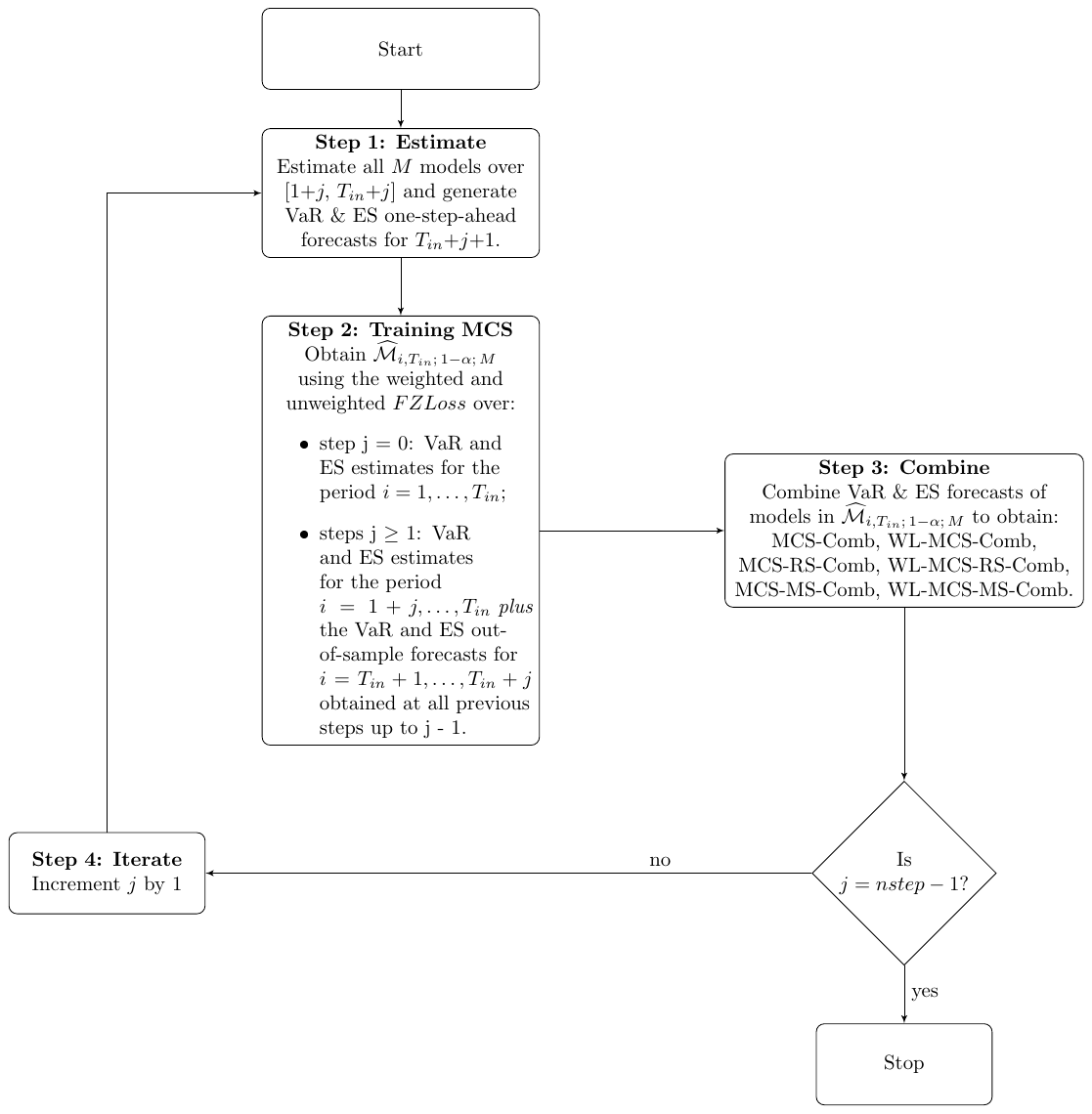}}
	\end{figure}

A summary diagram illustrating the previous algorithm is reported in Figure \ref{fig:summary_diagram}. We emphasize that in Step 2 of the algorithm the in-sample estimates are used solely at the beginning of the \emph{training} MCS procedures, while their influence gradually diminishes as they are replaced by out-of-sample forecasts. Accordingly, in the evaluation phase of the forecasts, we discard the initial part of the out-of-sample period (corresponding to the first $T_{in}/2$ observations), which constitutes a burn-in phase, to ensure that the combining weights of the proposed predictors are computed using at least 50\% out-of-sample forecasts and are increasingly driven by out-of-sample information.

%      
%      
%In addition to the \textcolor{red}{six} combined predictors, for the purpose of comparison, we also construct two standard benchmarks, denoted as {EW-Comb} and {Median-Comb}. In these benchmarks, all the VaR and ES forecasts from the $M$ models are jointly combined by taking their average and median, respectively. This approach has also been employed by \cite{mcaleer2013gfc} 
%for VaR forecasts, \textcolor{red}{while \cite{Taylor:2020} applied the EW-Comb method to combine VaR and ES forecasts. The set of combinations is then completed by the RS-Comb and MS-Comb predictors, where all the models are subject to the relative score and minimum score combination procedures, respectively, as in \cite{Taylor:2020}.}

\section{Empirical Analysis \label{sec:emp_analysis}}
We test our MCS-based combined predictors on time series of daily close-to-close log-returns for nine stock market indices: S\&P 500, Shanghai Composite, NASDAQ, Euro Stoxx 50, Nikkei, Hang Seng, BSESN, MXX, and Bovespa, observed over the period from January 2013 to June 2022. Together, the selected indices represent approximately 71.7\% of the total global market capitalization, based on exchange-level market capitalization figures (with S\&P 500 and Euro Stoxx 50 used as proxies for NYSE and Euronext, respectively), according to data from the \href{https://focus.world-exchanges.org/issue/june-2024/market-statistics}{\textcolor{blue}{World Federation of Exchanges}} (June 2024). Moreover, the selected indices ensure a broad geographical coverage, spanning major developed markets (United States, Euro area, Japan, and Hong Kong) and key emerging economies (China, India, Mexico, and Brazil), thereby capturing a large share of global financial activity.

Here we report the results for the S\&P 500 and Shanghai Composite indices at the 2.5\% coverage level, together with a summary table covering all indices at both the 2.5\% and 1\% coverage levels. The complete set of results is available on the supplementary material 
\href{https://vincenzocandila.shinyapps.io/comb_supp_mat/}{\textcolor{blue}{website}}. 

The additional exogenous variables used to enhance the information set consist of three high-frequency-based variables, observed on a daily basis and one low-frequency variable, observed monthly. The high-frequency variables are the realized volatility \citep{Andersen_et_al:2001} at 5 minutes (RVOL5), the realized bipower variation \citep{BarndorffNielsen:Shephard:2004} with subsampling (RB-SS) and the realized kernel \citep[RK,][]{BarndorffNielsen:Hansen:Lunde:Shephard:2008, BarndorffNielsen:Hansen:Lunde:Shephard:2009}. The low-frequency variable is the Economic Policy Uncertainty  \citep[EPU,][]{Baker_et_al:2016}. All the data have been collected from the Oxford-Man Realized library, except EPU which has been collected from the authors' \href{https://www.policyuncertainty.com/}{\textcolor{blue}{repository}}.
Table \ref{tab:sum_stat} illustrates the main summary statistics of the S\&P 500 and Shanghai Composite indices, for the period under investigation.

\begin{table}[htbp]
	\centering
		\caption{Summary statistics}
	\vspace{-0.35cm}
	\label{tab:sum_stat}
 \begin{adjustbox}{max height=0.7\textheight , max width=\textwidth}
  \begin{threeparttable}
\begin{tabular}{l  rrrrrrr}
\toprule
 &     {Obs.} &          {Min.} &         {Max.} &      {Mean} &     {SD} &       {Skew.} & {Kurt.}	\\
\midrule
 \multicolumn{8}{l}{\underline{S\&P 500; Sample period: from 2013-01-02 to 2022-06-28}}\\
log-returns & 2376 & -0.127 & 0.090 & 0.000 & 0.011 & -0.971 & 18.376 \\  
RVOL5 & 2376 & ~0.001 & 0.064 & 0.007 & 0.005 & 4.329 & 32.056 \\ 
RB-SS & 2376 & ~0.001 & 0.061 & 0.006 & 0.005 & 4.274 & 30.406 \\ 
RK & 2376 & ~0.001 & 0.058 & 0.007 & 0.005 & 3.933 & 26.599 \\ 

 \addlinespace
 \hdashline
 \addlinespace 
  \multicolumn{8}{l}{\underline{Shanghai Comp.; Sample period: from 2013-01-04 to 2022-06-28}}\\
log-returns & 2297 & -0.089 & 0.056 & 0.000 & 0.014 & -1.141 & 7.461 \\ 
RVOL5 & 2297 & ~0.002 & 0.064 & 0.009 & 0.006 & 3.418 & 17.413 \\ 
RB-SS & 2297 & ~0.002 & 0.067 & 0.009 & 0.006 & 3.509 & 18.768 \\ 
RK & 2297 & ~0.002 & 0.069 & 0.009 & 0.006 & 3.310 & 16.647 \\ 

 \addlinespace
 \hdashline
 \addlinespace 
%   \multicolumn{8}{l}{\underline{Euro Stoxx 50; Sample period: from 2013-01-02 to 2022-06-28}}\\
% log-returns & 2413 & -0.120 & 0.087 & 0.000 & 0.012 & -0.707 & 8.658 \\ 
%  RVOL5 & 2413 & ~0.000 & 0.074 & 0.009 & 0.006 & 4.096 & 30.780 \\ 
%RB-SS & 2413 & ~0.000 & 0.059 & 0.008 & 0.005 & 3.548 & 23.780 \\ 
% RK & 2413 & ~0.000 & 0.094 & 0.009 & 0.006 & 4.544 & 40.642 \\ 
% \addlinespace
% \hdashline
% \addlinespace 

  \multicolumn{8}{l}{\underline{EPU; Sample period: from 01-01-2010 to 01-06-2022}}\\
$\Delta{EPU}_t$ & 150 & -0.390 & 0.869 & 0.025 & 0.202 & 1.329 & 3.395\\
\bottomrule
\addlinespace
 \end{tabular}
 \vspace{-0.2cm}
 \begin{tablenotes}[flushleft]
   \setlength\labelsep{0pt}
   \footnotesize
\item \textbf{Notes}:  The table reports the number of observations (Obs.), the minimum (Min.) and maximum (Max.), the mean, standard deviation (SD), Skewness (Skew.) and excess Kurtosis (Kurt.). S\&P 500 and Shanghai Composite are obtained as the {close}-to-close log-returns. RVOL5 stands for the realized volatility \citep{Andersen_et_al:2001} at 5 minutes, RB-SS stands for the realized bipower variation \citep{BarndorffNielsen:Shephard:2004} with subsampling, and RK stands for the realized kernel \citep{BarndorffNielsen:Hansen:Lunde:Shephard:2008, BarndorffNielsen:Hansen:Lunde:Shephard:2009}. EPU stands for Economic Policy Uncertainty \citep{Baker_et_al:2016}, observed monthly, with $\Delta{EPU}_t= (EPU_t - EPU_{t-1} )/EPU_{t-1}$. 
\end{tablenotes}
\end{threeparttable}
\end{adjustbox}
\end{table} 

The chosen set of candidate models covers a wide range of frequently used parametric, semi-parametric and non-parametric techniques, as well as methods based on intraday data and mixed frequency variables. Table \ref{tab:mod_univ} reports the set of candidate models. Overall, we employ 18 parametric, one non-parametric model (with five window lengths for quantile estimation) and 9 semi-parametric models, for a total of 32 specifications.

	\begin{table}[htbp]
		\centering
		\caption{Model universe \label{tab:mod_univ}}
		\begin{adjustbox}{max width=1\textwidth, max height=0.45\textheight}
		  \begin{threeparttable}
			\begin{tabular}{l c c}
				\toprule  
				Model      &   Functional form     & Err. Distr.\\
				\midrule                                                                                                          
					\multirow{2}{*}{RM-N, RM-N-CF, RM-t (\textcolor{blue}{\citealt{Morgan:1996}})}              & $r_{i}|\mathcal{F}_{i-1}  = \sqrt{h_{i}} \eta_{i}$ & $\eta_{i}\overset{i.i.d}{\sim} \mathcal{N}\left(0, 1\right), \, \eta_{i}\overset{i.i.d}{\sim} \mathcal{N}\left(0, 1\right) + CF, \, \eta_{i}\overset{i.i.d}{\sim} t_{\nu}$ \\  
				&      $h_{i} = \zeta h_{i-1} +  (1-\zeta) r_{i-1}^2$& \\			
		\addlinespace
				\hdashline
				\addlinespace				
		\multirow{2}{*}{GARCH-N, GARCH-N-CF, GARCH-t (\textcolor{blue}{\citealt{Bollerslev:1986}})}              & $r_{i}|\mathcal{F}_{i-1}  = \sqrt{h_{i}} \eta_{i}$ & $\eta_{i}\overset{i.i.d}{\sim} \mathcal{N}\left(0, 1\right), \, \eta_{i}\overset{i.i.d}{\sim} \mathcal{N}\left(0, 1\right) + CF, \, \eta_{i}\overset{i.i.d}{\sim} t_{\nu}$\\  
				&      $h_{i} = \omega + \alpha_{}  r_{i-1}^2 + \beta h_{i-1}$& \\
				\addlinespace
				\hdashline
				\addlinespace
				\multirow{2}{*}{GJR-N, GJR-N--CF, GJR-t (\textcolor{blue}{\citealt{Glosten:Jaganathan:Runkle:1993}})}              & $r_{i}|\mathcal{F}_{i-1}  = \sqrt{h_{i}} \eta_{i}$ & $\eta_{i}\overset{i.i.d}{\sim} \mathcal{N}\left(0, 1\right), \, \eta_{i}\overset{i.i.d}{\sim} \mathcal{N}\left(0, 1\right) + CF, \, \eta_{i}\overset{i.i.d}{\sim} t_{\nu}$\\  
				&      $h_{i} = \omega + \left(\alpha_{} + \gamma_{} \mathbbm{1}_{\left( r_{i-1} < 0 \right)} \right) r_{i-1}^2 + \beta_{} h_{i-1}$& \\
				\addlinespace
				\hdashline
				\addlinespace  
				
				\multirow{3}{*}{RGARCH-N, RGARCH-N-CF, RGARCH-t (\textcolor{blue}{\citealt{Hansen:Huang:Shek:2012}})}			& $r_{i}|\mathcal{F}_{i-1}  = \sqrt{h_{i}} \eta_{i}$ & $\eta_{i}\overset{i.i.d}{\sim} \mathcal{N}\left(0, 1\right), \, \eta_{i}\overset{i.i.d}{\sim} \mathcal{N}\left(0, 1\right) + CF, \, \eta_{i}\overset{i.i.d}{\sim} t_{\nu}$\\	
				& 	$h_{i}=const + \beta h_{i-1} + \alpha x_{i-1}$& \\   
				&	$x_{i}=const_{x} + \delta h_{i} + \tau_1 \eta_{i} + \tau_2 \left(\eta_{i}^2-1\right) + \sigma_u u_{i}$  &\\
%					\addlinespace
%				\hdashline
%				\addlinespace
%				\multirow{2}{*}{HAR--N, HAR--t (\textcolor{blue}{\citealt{Corsi:2009}})} 	& $r_{i}|\mathcal{F}_{i-1}  = \sqrt{\tilde{h}_{i}} \eta_{i}$ & $\eta_{i}\overset{i.i.d}{\sim} \mathcal{N}\left(0, 1\right), \, \eta_{i}\overset{i.i.d}{\sim} t_{\nu}$\\			&$\tilde{h}_{i}= const+ \beta_1 {\tilde{h}}_{i-1}+\beta_5 \overline{\tilde{h}}_{(i-1):(i-5)}+\beta_{22} \overline{\tilde{h}}_{(i-1):(i-22)} + \eta_{i}$ & \\	
		
					\addlinespace
				\hdashline
				\addlinespace   
				\multirow{2}{*}{HS-w (\textcolor{blue}{\citealt{Hendricks:1996}})} & $VaR_{i}(\tau)={Q}_{\bm r_{i}^{w}}(\tau)$\\ 
				&$\bm r_{i}^{w} = (r_{i-w},r_{i-w+1},\dots,r_{i-1})$    \\ 
				\addlinespace
				\hdashline
				\addlinespace    
				
				CAViaR-SAV  (\textcolor{blue}{\citealt{Engle:Manganelli:2004}})     &     $VaR_{i}(\tau) = \beta_0+ \beta_1 VaR_{i-1}(\tau)+ \beta_2 |r_{i-1}| $&\\		
				\addlinespace
				\hdashline
				\addlinespace
				\multirow{2}{*}{CAViaR-AS  (\textcolor{blue}{\citealt{Engle:Manganelli:2004}})}      &    $VaR_{i}(\tau) =  \beta_0+  \beta_1 VaR_{i-1}(\tau)+ (\beta_{2}\mathbbm{1}_{(r_{i-1}>0)}+$&\\
				& $\beta_{3} \mathbbm{1}_{(r_{i-1}<0)}) |r_{i-1}|$ &\\
				\addlinespace
				\hdashline
				\addlinespace
				CAViaR-IG   (\textcolor{blue}{\citealt{Engle:Manganelli:2004}})    &    $VaR_{i}(\tau) =  -\sqrt {\beta_0+ \beta_1 VaR_{i-1}^2(\tau)+ \beta_2 r_{i-1}^2 }$&\\ 
				\addlinespace
				\hdashline
				\addlinespace
				CAViaR--X    (\textcolor{blue}{\citealt{Gerlach:Wang:2020}})   &     $VaR_{i}(\tau) = \beta_0+ \beta_1 VaR_{i-1}(\tau)+ \beta_2 x_{i-1} $&\\      
		\addlinespace
				\hdashline
				\addlinespace				
    \multirow{4}{*}{MF-X    (\textcolor{blue}{\citealt{candila2023mixed}})}  & $r_{i,t}|\mathcal{F}_{i-1,t}  =\sqrt{h_{i,t}}  \eta_{i,t}$ & $\eta_{i,t}\overset{i.i.d}{\sim} \left(0, 1\right)$\\         
 				        &   $\sqrt{h_{i,t}}=  (\beta_0 + \theta |WS_{t-1}|+ \beta_1 |r_{i-1,t}| + \ldots + $ &\\
 				        & $\beta_q |r_{i-q,t}| + \beta_X |X_{i-1,t}|)$&\\
        & $WS_{t-1}= \sum_{k=1}^K \delta_k(\omega)MV_{t-k}$&\\        			
		
				\bottomrule
			\end{tabular}
\begin{tablenotes}[flushleft]
   \setlength\labelsep{0pt}
   \footnotesize
\item \textbf{Notes}:  The table reports the functional forms and references of the models used in this work. \textit{RM} stands for RiskMetrics, \textit{RGARCH} for Realized GARCH, \textit{MF} for Mixed-Frequency, with $MV$ denoting a low-frequency Macro-Variable. The column \textit{Err. Distr.} shows the error distributions for the parametric models: Normal and Student-\(t\) with \(\nu\) degrees of freedom. The parametric models also employ the Cornish–Fisher (CF) expansion to compute VaR and ES forecasts. All semi-parametric models estimate ES following the approach of \cite{taylor2019forecasting} and \cite{Gerlach:Wang:2020}, which allows for the joint estimation of VaR and ES. The ES is obtained using the multiplicative VaR specification, that is,
$ES_{i}(\tau) = \bigl( 1 + \exp(\gamma_0) \bigr) \cdot \text{VaR}_{i}(\tau)$, where $\gamma_0$ is estimated by maximizing the AL likelihood. In the presentation of the MF-X model, we adopt the double time index, $i,t$, where $t=1,\ldots,T$ denotes a low-frequency time period (for instance, monthly, quarterly, and so forth) and $i=1,\ldots,N_t$ refers to the day of the low-frequency period, with $N_t$ representing a varying number of days in $t$, and an overall number $N$ of daily observations $N=\sum_{t=1}^T N_t$. 
\end{tablenotes}
\end{threeparttable} 
\end{adjustbox}
\end{table}

Semi-parametric models are estimated following the methodological approach developed by \cite{taylor2019forecasting} and \cite{Gerlach:Wang:2020}. Specifically, the underlying theoretical framework exploits equivalence between the conventional quantile regression estimator and the maximum likelihood estimator derived from the AL distribution \citep{koenker1999goodness}. \cite{taylor2019forecasting} extends this theoretical result by incorporating the ES measure into the likelihood specification, thus enabling the simultaneous estimation of models for both VaR and ES, under the assumption of zero-mean return innovations.

Throughout the analysis, we use a rolling window of $T_{in}=1000$ daily observations, which moves forward by one day at each step. This implies that all models are re-estimated on a daily basis. As concerns the exponential decay parameter of the $FZLoss$ in \eqref{eq:wfzloss}, we fix $\lambda=0.06$. %\textcolor{red}{The value of $\lambda = 0.06$ implies that the decay time is approximately equivalent to four trading months \citep[][sec. 5.3.1.2]{Morgan:1996}. In other words, the performance of each predictor is computed as an exponentially decaying weighted average of the loss values observed over approximately the last four trading months.}
The significance level of the $training$ MCS is set to $\alpha = 0.25$, while a robustness analysis for different values of $\lambda$ and $\alpha$ is provided in the supplementary material \href{https://vincenzocandila.shinyapps.io/comb_supp_mat/}{\textcolor{blue}{website}}.  

The first part of our analysis focuses on evaluating the performance of all models on a period-by-period basis. As mentioned previously, as time progresses, each training period incorporates an increasing number of out-of-sample forecasts.

%In these training periods, some models perform well in some periods, some others not. In other words, there is no clear evidence of a single model or subset of models consistently outperforming the others, regardless of the backtests employed. 

\begin{figure}[htbp]
%\thisfloatpagestyle{empty}
\vspace*{-1.2cm}
	\centering
	\caption{Shanghai Composite. Backtests and MCS over the training periods} 
	\label{fig:backtests_mcs_training}
	\vspace{-0.35cm}
	\begin{subfigure}[b]{0.8\textwidth}
		\centering
		\includegraphics[width=1\linewidth]{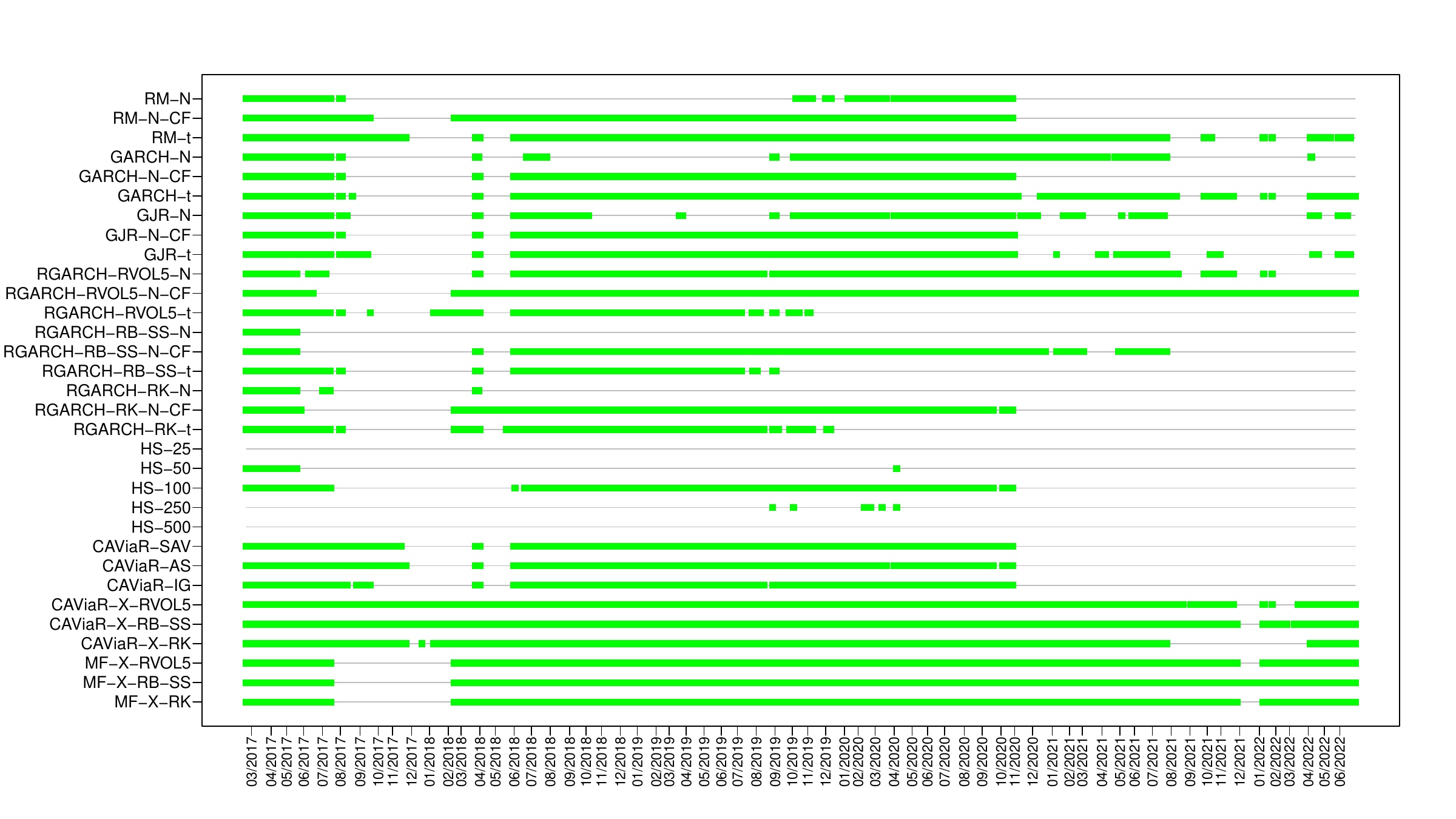}
		\caption{Backtests \label{fig:shanghai_backtest}}
	\end{subfigure}%

\vspace{-0.5cm}   % molto meno spazio

		\begin{subfigure}[b]{0.8\textwidth}
		\centering
		\includegraphics[width=1\linewidth]{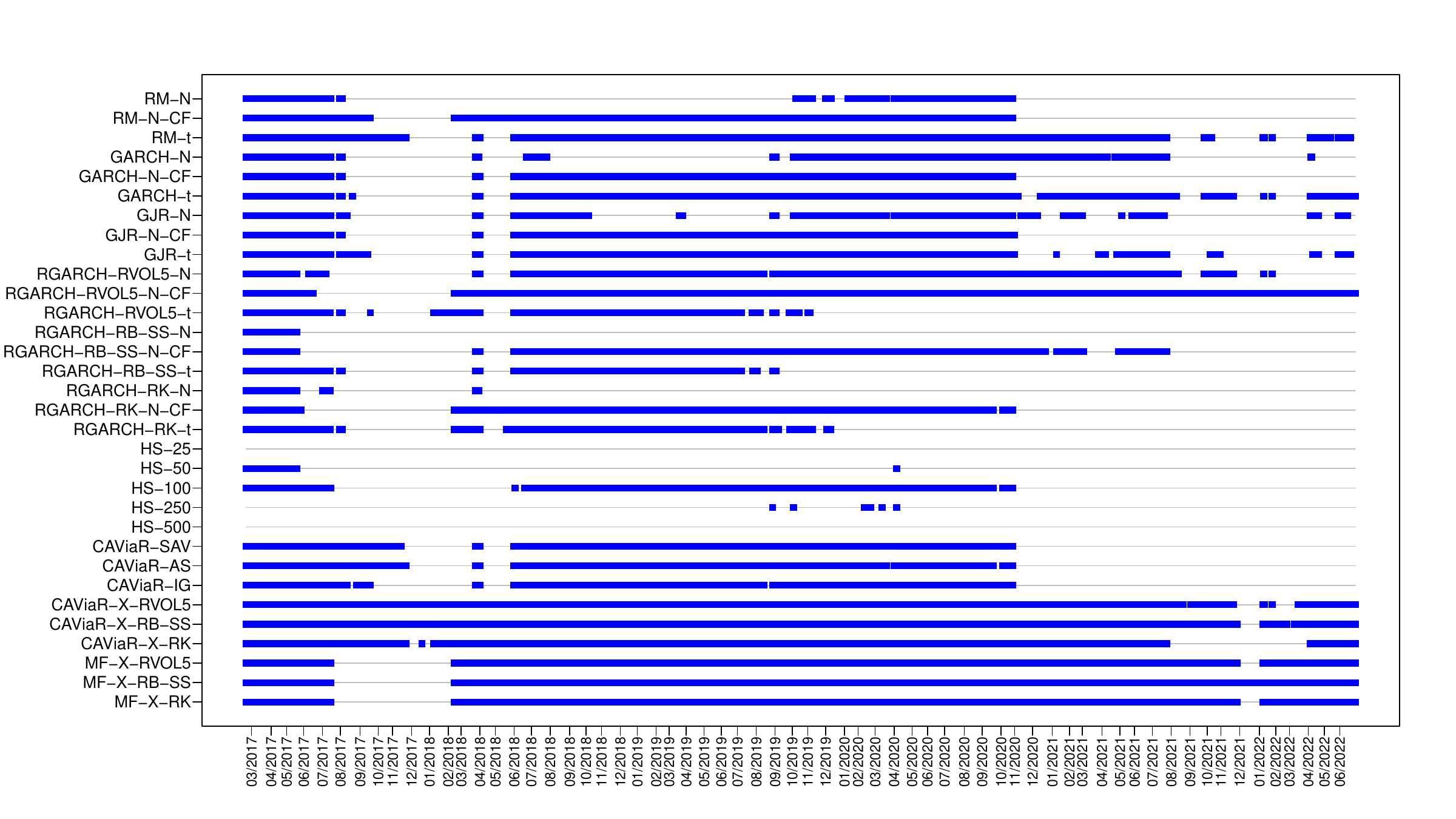}
		\caption{MCS with the unweighted $FZLoss$\label{fig:shanghai_mcs_unwei}}
	\end{subfigure}

\vspace{-0.5cm}   % molto meno spazio
		\begin{subfigure}[b]{0.8\textwidth}
		\centering
		\includegraphics[width=1\linewidth]{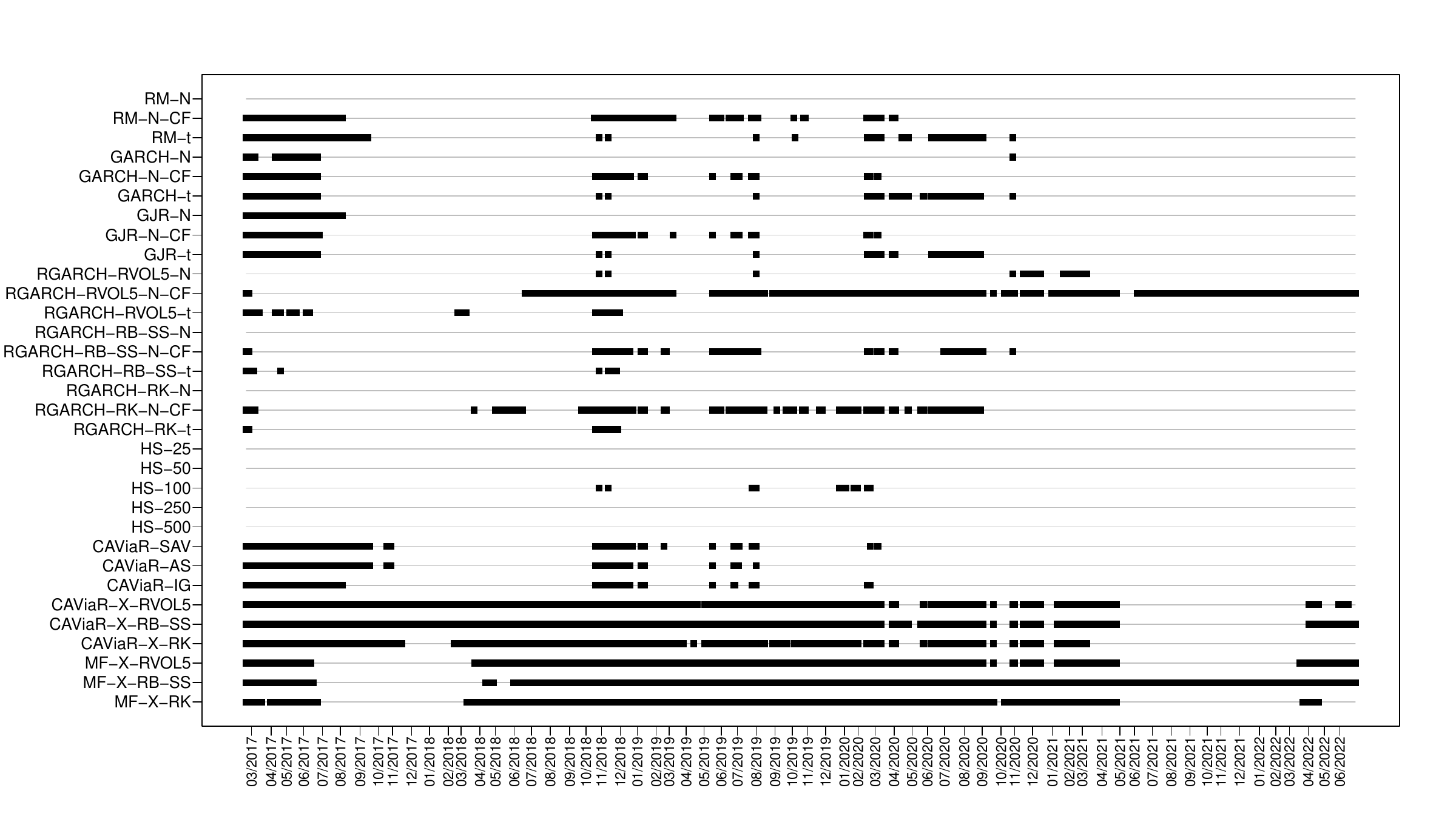}
		\caption{MCS with the weighted $FZLoss$\label{fig:shanghai_mcs_wei}}
	\end{subfigure}
	%\hfill
				\vspace{-0.1cm}
	\begin{minipage}[t]{1\textwidth}
		\textbf{Notes:} Plots of models passing all six backtesting procedures reported in Table~\ref{tab:backtests} (top panel, green squares), and belonging to the MCS according to the \emph{unweighted} (middle panel, blue squares) and \emph{weighted} (bottom panel, black squares) $FZLoss$, over the training periods.
	\end{minipage}
	
\end{figure}

The top panel of Figure \ref{fig:backtests_mcs_training}, dedicated to the Shanghai Composite Index, displays a green square when a model successfully passes all six backtests employed in this study (i.e., UC, CC, DQ, BD-1, BD-2 and BD-3, synthetically described in Table \ref{tab:backtests}). The passing of these backtests provides strong evidence of adequate VaR and ES measures. Notably, no single model consistently passes all the backtests over time. 

When attention is on the SSM of the MCS, the situation appears similar. The central and bottom panels of Figure \ref{fig:backtests_mcs_training} show the SSM composition, where a square indicates that the model belongs to the set, computed, according to the unweighted (Figure \ref{fig:shanghai_mcs_unwei}) and weighted (Figure \ref{fig:shanghai_mcs_wei}) $FZLoss$, respectively. 
Once again, no single model consistently enters the SSM across all training periods.

It is worth noting that using \emph{weighted} $FZLoss$ leads to greater sparsity, with many fewer models entering the SSM compared to the \emph{unweighted} $FZLoss$. 

Overall, based on the backtesting and MCS procedures in the training periods, it can be concluded that nonparametric specifications do not perform adequately well, parametric models only occasionally yield satisfactory results, whereas semi-parametric models generally exhibit the best performance. These results are consistent across all indices, as can be seen in the supplementary material \href{https://vincenzocandila.shinyapps.io/comb_supp_mat/}{\textcolor{blue}{website}}.

	\begin{table}[htbp]
	\centering
	\caption{Backtests \label{tab:backtests}}
		\resizebox{1\textwidth}{!}{%
			\begin{threeparttable}
				\begin{tabular}{lrl}
					\toprule
			Label	& Name  & Null Hypothesis\\
					\midrule
UC & Unconditional Coverage \citep{kupiec1995techniques} & Correct number of VaR violations\\
   \addlinespace
  \hdashline
   \addlinespace 
   \multirow{2}{*}{CC} & \multirow{2}{*}{Conditional Coverage \citep{Christoffersen:1998}} & Correct number and independence of \\
	&&VaR violations	\\
	   \addlinespace
  \hdashline
   \addlinespace 
   \multirow{2}{*}{DQ} & \multirow{2}{*}{Conditional Coverage \citep{Engle:Manganelli:2004}} & Correct number and independence of \\
	&&VaR violations	\\
		   \addlinespace
  \hdashline
   \addlinespace 
BD-1 & Strict ES Regression \citep{Bayer:Dimitriadis:2020} &  $\beta_0=0$ and $\beta_1=1$ in Regression 1\\
   \addlinespace
  \hdashline
   \addlinespace 
BD-2 & Auxiliary ES Regression \citep{Bayer:Dimitriadis:2020} & $\beta_0=0$ and $\beta_1=1$ in Regression 2\\
		   \addlinespace
  \hdashline
   \addlinespace 
BD-3 & Strict Intercept \citep{Bayer:Dimitriadis:2020} & $\beta_0=0$ in Regression 3 \\
	\bottomrule
				\end{tabular}
				\begin{tablenotes}[flushleft]
					\setlength\labelsep{0pt}
\item  \textbf{Notes}:  Regression 1: $r_{i}= \beta_0 + \beta_1 ES_{i}(\tau) + u_{i}$. Regression 2: $r_{i}= \beta_0 + \beta_1 ES_{i}(\tau) + \beta_2 VaR_{i}(\tau)+ u_{i}$.  Regression 3: $e_{i}= \beta_0  + u_{i}$, with  $e_{i}=r_{i} - ES_{i}(\tau)$.
				\end{tablenotes}
			\end{threeparttable}
		}
	\end{table}

In the subsequent part of our analysis and motivated by the findings in Figure \ref{fig:backtests_mcs_training}, we construct the six proposed predictors: MCS-Comb, WL-MCS-Comb, MCS-RS-Comb, WL-MCS-RS-Comb, MCS-MS-Comb and WL-MCS-MS-Comb, as explained above. These predictors are developed by combining the out-of-sample forecasts of VaR and ES from the best-performing models identified adaptively during the \emph{training} periods. As benchmarks, we also consider four additional combined predictors: the simple average of all risk forecasts in the model universe (EW-Comb); their median (Median-Comb); and the combined predictors obtained by applying the RS-Comb and MS-Comb approaches to all forecasts (RS-Comb and MS-Comb, respectively). A comprehensive assessment of the combined predictors, together with the benchmark predictors EW-Comb, Median-Comb, RS-Comb and MS-Comb, as well as the entire model universe, is presented in Tables \ref{tab:sp500_oos_evaluation} (S\&P 500) and \ref{tab:shanghai_oos_evaluation} (Shanghai Composite Index). The evaluation relies on the six backtests and the MCS procedure over the whole out-of-sample period, discarding the first 500 observations (i.e., $T_{in}/2$), which correspond to a burn-in phase, as discussed above. In Tables \ref{tab:sp500_oos_evaluation} and  \ref{tab:shanghai_oos_evaluation}, light shades of gray indicate success in the backtesting procedures, at significance level of $\alpha=0.05$, while dark shades of gray indicate inclusion in the MCS, at significance level $\alpha=0.25$. Some important points can be highlighted. First, CAViaR-X models generally pass all backtesting procedures, while parametric and non-parametric models do not perform adequately, except for a few cases. Second, the benchmark combined predictor EW-Comb passes all six backtests for the S\&P 500 but not for the Shanghai Composite Index. Remarkably, the other two benchmarks, RS- and MS-Comb, pass all six backtests only for the S\&P 500. Third and more importantly, all the proposed and MCS-based combined predictors pass all the backtests, regardless of the index considered, except WL-MCS-Comb for the S\&P 500 and MCS-RS-Comb for the Shanghai Composite Index. Fourth, the SSM for the S\&P 500 includes nine models, namely the three CAViaR-X specifications, RS-Comb, and all proposed predictors except MCS-MS-Comb, whereas the SSM for the Shanghai Composite Index is smaller, comprising one RGARCH model and one CAViaR-X specification (using RB-SS), along with three combined predictors: MCS-RS-Comb, WL-MCS-RS-Comb, and WL-MCS-MS-Comb. These results are generally robust to small variations in the parameter $\lambda$ of Eq.~\eqref{eq:wfzloss}, as can be seen in the supplementary material \href{https://vincenzocandila.shinyapps.io/comb_supp_mat/}{\textcolor{blue}{website}}. For instance, the SSM of the Shanghai Composite Index computed using $\lambda=0.03$ instead of $\lambda=0.06$ comprises the same models as those obtained with $\lambda=0.06$, except for WL-MCS-RS-Comb, whereas when $\lambda=0.09$ the SSM coincides with that obtained under $\lambda=0.06$, except for WL-MCS-MS-Comb.

A clear picture in favor of the proposed combined predictors emerges from Table \ref{tab:overall_oos_evaluation}, which summarizes successful backtests and MCS inclusions across all nine indices and the two coverage levels. In this table, we report the number of times each model passes all backtests (first column for $\tau=0.025$ and fourth column for $\tau=0.01$, with the label \virg{\#BT}), belongs to the MCS (second column for $\tau=0.025$ and fifth column for $\tau=0.01$, with the label \virg{\#MCS}), and satisfies both criteria (third column for $\tau=0.025$ and sixth column for $\tau=0.01$, with the label \virg{\#BT and \#MCS}). Higher values indicate better performance, with 9 (reported in bold in the table) being the maximum possible, with dark green cells highlighting the best performance in each column and light green cells the second-best. Notably, on the MCS criterion, three of the proposed predictors, namely MCS-RS-Comb, WL-MCS-RS-Comb and WL-MCS-MS-Comb, achieve the maximum score of 9 at $\tau=0.025$, meaning they belong to the SSM across all nine indices without exception, a result matched by no individual model or benchmark combination. With respect to the backtesting criterion, WL-MCS-RS-Comb exhibits a \#BT count of 7 across both levels of coverage, performing comparably to the most successful individual models or benchmark combination. Most importantly, on the joint criterion, the WL-MCS-RS-Comb predictor stands out as the overall best-performing specification, recording the highest joint count (\#BT and \#MCS = 7) at $\tau=0.025$ and a joint count of 6 at $\tau=0.01$, the highest observed across all models at either coverage level. 
By contrast, the benchmark combined predictors EW-Comb and Median-Comb reach at most a joint count of 3, while the global RS-Comb and MS-Comb, despite competitive MCS inclusion rates, are consistently dominated by the proposed MCS-based predictors on the joint criterion. These findings strongly support the adoption of the proposed MCS-based combination framework as a robust approach to tail-risk forecasting across heterogeneous market environments.

It is worth noting that the CAViaR-X models represent the strongest individual competitors, particularly at
$\tau=0.01$. However, no single CAViaR-X
specification dominates uniformly across coverage levels or evaluation criteria, as their relative performance is sensitive to the choice of realized volatility measure employed. 
This measure-dependence represents an additional source of model uncertainty that the proposed combination strategies are specifically designed to mitigate.

In summary, the proposed MCS-based combined predictors exhibit robust and consistent performance across indices and risk levels, as measured by both backtesting and MCS inclusion criteria. Among them, WL-MCS-RS-Comb emerges as the overall best-performing specification, achieving the highest joint counts at
both $\tau=0.025$ and $\tau=0.01$ and consistently outperforming the benchmark combinations. These findings confirm that leveraging the MCS within a dedicated tail-risk combination framework provides an effective approach to mitigating the multiple sources of uncertainty that may compromise the accuracy of individual VaR and ES forecasts.

%Only two models, the CAViaR-X-RVOL5 and CAViaR-X-RB-SS, together with the proposed combined predictor WL-MCS-Comb, reach the maximum value of 6. While CAViaR-X-RVOL5 achieves the best performance at $\tau = 0.025$, CAViaR-X-RB-SS emerges as the best model at $\tau = 0.01$. By contrast, WL-MCS-Comb exhibits consistently strong performance at both coverage levels. Moreover, the proposed WL-MCS-Comb consistently outperforms the benchmarks: EW-Comb, Median-Comb and RS- and MS-Comb. 
%
%In summary, the proposed combined predictors, with WL-MCS-Comb yielding the best performance, exhibit robust results in both backtests and MCS. Overall, adopting the proposed combined predictors proves effective in mitigating the impact of various sources of uncertainty that may affect VaR and ES forecasts.

	\begin{table}[htbp]
	\centering
	\caption{S\&P 500 out-of-sample evaluation \label{tab:sp500_oos_evaluation}}
		\resizebox{0.80\textwidth}{!}{%
			\begin{threeparttable}
				\begin{tabular}{lrrrrrrr}
					\toprule
				& UC & CC & DQ & BD-1 & BD-2 & BD-3 & $FZLoss$ \\
					\midrule
RM-N & 0.00 & 0.00 & 0.00 & 0.00 & 0.00 & 0.00 & -3.182 \\ 
RM-N-CF & 0.68 & 0.71 & 0.07 & 0.00 & 0.00 & 0.02 & -3.330 \\ 
RM-t & 0.00 & 0.00 & 0.00 & 0.12 & 0.18 & 0.29 & -3.258 \\ 
GARCH-N & 0.01 & 0.03 & 0.13 & 0.00 & 0.01 & 0.01 & -3.348 \\ 
GARCH-N-CF & 0.38 & 0.47 & 0.84 & 0.02 & 0.01 & 0.04 & -3.457 \\ 
GARCH-t & 0.02 & 0.08 & 0.22 & 0.21 & 0.22 & 0.68 & -3.401 \\ 
GJR-N & 0.01 & 0.05 & 0.04 & 0.02 & 0.02 & 0.12 & -3.336 \\ 
\cellcolor{gray!25}GJR-N-CF & 0.18 & 0.30 & 0.69 & 0.13 & 0.08 & 0.09 & -3.426 \\ 
\cellcolor{gray!25}GJR-t & 0.06 & 0.18 & 0.20 & 0.15 & 0.12 & 0.82 & -3.398 \\ 
RGARCH-RVOL5-N & 0.00 & 0.00 & 0.00 & 1.00 & 1.00 & 0.12 & -3.200 \\ 
RGARCH-RVOL5-N-CF & 0.00 & 0.00 & 0.00 & 0.65 & 0.62 & 0.93 & -3.510 \\ 
RGARCH-RVOL5-t & 0.00 & 0.00 & 0.00 & 0.05 & 1.00 & 0.74 & -3.239 \\ 
RGARCH-RB-SS-N & 0.00 & 0.00 & 0.00 & 1.00 & 0.01 & 1.00 & -3.106 \\ 
RGARCH-RB-SS-N-CF & 0.00 & 0.00 & 0.00 & 0.04 & 0.05 & 0.02 & -3.490 \\ 
RGARCH-RB-SS-t & 0.00 & 0.00 & 0.00 & 0.00 & 0.05 & 0.00 & -3.141 \\ 
RGARCH-RK-N & 0.00 & 0.00 & 0.00 & 1.00 & 1.00 & 1.00 & -2.841 \\ 
RGARCH-RK-N-CF & 0.00 & 0.00 & 0.00 & 1.00 & 0.39 & 1.00 & -3.350 \\ 
RGARCH-RK-t & 0.00 & 0.00 & 0.00 & 0.00 & 0.00 & 0.00 & -2.841 \\ 
\addlinespace
\hdashline
\addlinespace
HS-25 & 0.00 & 0.00 & 0.00 & 0.00 & 0.00 & 0.00 & -2.583 \\ 
HS-50 & 0.00 & 0.00 & 0.00 & 0.00 & 0.00 & 0.01 & -3.017 \\ 
HS-100 & 0.00 & 0.02 & 0.00 & 0.02 & 0.02 & 0.30 & -3.103 \\ 
HS-250 & 0.00 & 0.01 & 0.00 & 0.02 & 0.03 & 0.40 & -2.915 \\ 
HS-500 & 0.04 & 0.01 & 0.00 & 0.00 & 0.00 & 0.75 & -2.844 \\ 
\addlinespace
\hdashline
\addlinespace
\cellcolor{gray!25}SAV & 0.29 & 0.56 & 0.48 & 0.89 & 0.72 & 0.55 & -3.463 \\ 
AS & 0.00 & 0.00 & 0.00 & 0.18 & 0.21 & 0.95 & -3.283 \\ 
\cellcolor{gray!25}IG & 0.10 & 0.09 & 0.45 & 0.67 & 0.51 & 0.48 & -3.468 \\ 
\cellcolor{gray!25}CAViaR-X-RVOL5 & 0.81 & 0.87 & 0.97 & 0.77 & 0.90 & 0.70 & \cellcolor{gray!75}-3.603 \\ 
\cellcolor{gray!25}CAViaR-X-RB-SS & 0.65 & 0.83 & 0.59 & 0.58 & 0.76 & 0.44 & \cellcolor{gray!75}-3.635 \\ 
\cellcolor{gray!25}CAViaR-X-RK & 0.20 & 0.45 & 0.66 & 0.89 & 0.95 & 0.66 & \cellcolor{gray!75}-3.595 \\ 
MF-X-RVOL5 & 0.29 & 0.24 & 0.26 & 0.00 & 0.03 & 0.26 & -3.472 \\ 
MF-X-RB-SS & 0.20 & 0.18 & 0.38 & 0.00 & 0.00 & 0.15 & -3.472 \\ 
MF-X-RK & 0.00 & 0.01 & 0.00 & 0.00 & 0.07 & 0.05 & -3.460 \\ 
\addlinespace
\hdashline
\addlinespace
\cellcolor{gray!25}EW-Comb & 0.10 & 0.25 & 0.09 & 0.95 & 0.96 & 0.94 & -3.524 \\ 
Median-Comb & 0.00 & 0.02 & 0.00 & 0.73 & 0.64 & 0.44 & -3.509 \\ 
\addlinespace
\hdashline
\addlinespace
\cellcolor{gray!25}RS-Comb & 0.51 & 0.76 & 0.08 & 0.85 & 0.94 & 0.60 & \cellcolor{gray!75}-3.614 \\ 
\cellcolor{gray!25}MS-Comb & 0.06 & 0.18 & 0.41 & 0.89 & 0.80 & 0.92 & -3.517 \\ 
\addlinespace
\hdashline
\addlinespace
\cellcolor{gray!25}MCS-Comb & 0.10 & 0.25 & 0.06 & 0.98 & 0.97 & 0.90 & \cellcolor{gray!75}-3.578 \\ 
WL-MCS-Comb & 0.20 & 0.45 & 0.00 & 0.96 & 0.99 & 0.93 & \cellcolor{gray!75}-3.592 \\ 
\cellcolor{gray!25}MCS-RS-Comb & 0.39 & 0.67 & 0.09 & 0.84 & 0.95 & 0.63 & \cellcolor{gray!75}-3.609 \\ 
\cellcolor{gray!25}WL-MCS-RS-Comb & 0.39 & 0.67 & 0.09 & 0.86 & 0.95 & 0.62 & \cellcolor{gray!75}-3.609 \\ 
\cellcolor{gray!25}MCS-MS-Comb & 0.10 & 0.25 & 0.06 & 0.97 & 0.95 & 0.95 & -3.571 \\ 
\cellcolor{gray!25}WL-MCS-MS-Comb & 0.29 & 0.56 & 0.10 & 0.95 & 1.00 & 0.77 & \cellcolor{gray!75}-3.597 \\ 
 
		\bottomrule
				\end{tabular}
				\begin{tablenotes}[flushleft]
					\setlength\labelsep{0pt}
\item  \textbf{Notes}: Columns from UC to BD-3 report the p-values of the six backtesting procedures described in Table \ref{tab:backtests}. Column $FZLoss$ represents the averages of the $FZLoss$. Light shades of gray denote the success in the backtesting procedures, at significance level $\alpha = 0.05$. Dark shades of gray denote the inclusion in the MCS, at significance level $\alpha = 0.25$. Sample period: 2018-12-18 to 2022-06-28 (876 observations). VaR and ES are calculated at the level $\tau=0.025$. 
				\end{tablenotes}
			\end{threeparttable}
		}
	\end{table}
 
	\begin{table}[htbp]
	\centering
	\caption{Shanghai Composite out-of-sample evaluation \label{tab:shanghai_oos_evaluation}}
		\resizebox{0.80\textwidth}{!}{%
			\begin{threeparttable}
				\begin{tabular}{lrrrrrrr}
					\toprule
			 & UC & CC & DQ & BD-1 & BD-2 & BD-3 & $FZLoss$ \\
					\midrule
RM-N & 0.13 & 0.06 & 0.00 & 0.13 & 0.11 & 0.07 & -3.261 \\ 
RM-N-CF & 0.03 & 0.00 & 0.00 & 0.02 & 0.02 & 0.18 & -3.282 \\ 
RM-t & 0.19 & 0.07 & 0.00 & 0.49 & 0.47 & 0.43 & -3.321 \\ 
GARCH-N & 0.05 & 0.04 & 0.00 & 0.12 & 0.18 & 0.09 & -3.312 \\ 
GARCH-N-CF & 0.01 & 0.01 & 0.05 & 0.21 & 0.17 & 0.15 & -3.328 \\ 
GARCH-t & 0.64 & 0.06 & 0.00 & 0.75 & 0.78 & 0.58 & -3.350 \\ 
GJR-N & 0.08 & 0.14 & 0.03 & 0.14 & 0.15 & 0.08 & -3.323 \\ 
\cellcolor{gray!25}GJR-N-CF & 0.24 & 0.28 & 0.43 & 0.37 & 0.29 & 0.18 & -3.348 \\ 
GJR-t & 0.27 & 0.07 & 0.00 & 0.71 & 0.83 & 0.55 & -3.364 \\ 
RGARCH-RVOL5-N & 0.00 & 0.00 & 0.00 & 0.00 & 0.00 & 0.00 & -3.278 \\ 
RGARCH-RVOL5-N-CF & 0.03 & 0.01 & 0.00 & 0.66 & 0.66 & 0.56 & \cellcolor{gray!75}-3.486 \\ 
RGARCH-RVOL5-t & 0.00 & 0.00 & 0.00 & 0.00 & 0.00 & 0.00 & -2.642 \\ 
RGARCH-RB-SS-N & 0.00 & 0.00 & 0.00 & 0.00 & 0.00 & 0.00 & -2.179 \\ 
RGARCH-RB-SS-N-CF & 0.00 & 0.00 & 0.00 & 0.01 & 0.01 & 0.00 & -3.059 \\ 
RGARCH-RB-SS-t & 0.00 & 0.00 & 0.00 & 0.00 & 0.00 & 0.00 & -2.462 \\ 
RGARCH-RK-N & 0.00 & 0.00 & 0.00 & 0.00 & 0.00 & 0.00 & -1.402 \\ 
RGARCH-RK-N-CF & 0.00 & 0.00 & 0.00 & 0.00 & 0.00 & 0.00 & -2.659 \\ 
RGARCH-RK-t & 0.00 & 0.00 & 0.00 & 0.00 & 0.00 & 0.00 & -2.347 \\ 
\addlinespace
\hdashline
\addlinespace
HS-25 & 0.00 & 0.00 & 0.00 & 0.00 & 0.00 & 0.13 & -2.816 \\ 
HS-50 & 0.00 & 0.00 & 0.00 & 0.03 & 0.03 & 0.19 & -3.011 \\ 
HS-100 & 0.03 & 0.03 & 0.00 & 0.06 & 0.04 & 0.41 & -3.126 \\ 
HS-250 & 0.27 & 0.07 & 0.00 & 0.21 & 0.12 & 0.63 & -3.226 \\ 
HS-500 & 0.37 & 0.07 & 0.00 & 0.22 & 0.53 & 0.94 & -3.253 \\ 
\addlinespace
\hdashline
\addlinespace
\cellcolor{gray!25}SAV & 0.99 & 0.25 & 0.07 & 0.76 & 0.75 & 0.77 & -3.334 \\ 
AS & 0.64 & 0.06 & 0.00 & 0.20 & 0.35 & 0.95 & -3.247 \\ 
\cellcolor{gray!25}IG & 0.64 & 0.30 & 0.12 & 0.83 & 0.87 & 0.69 & -3.368 \\ 
\cellcolor{gray!25}CAViaR-X-RVOL5 & 0.99 & 0.82 & 0.56 & 0.93 & 0.90 & 0.86 & -3.444 \\ 
\cellcolor{gray!25}CAViaR-X-RB-SS & 0.64 & 0.30 & 0.13 & 0.99 & 1.00 & 0.96 & \cellcolor{gray!75}-3.473 \\ 
\cellcolor{gray!25}CAViaR-X-RK & 0.64 & 0.80 & 0.48 & 0.99 & 0.99 & 0.97 & -3.443 \\ 
\cellcolor{gray!25}MF-X-RVOL5 & 0.66 & 0.65 & 0.86 & 0.97 & 0.82 & 0.89 & -3.447 \\ 
MF-X-RB-SS & 0.36 & 0.07 & 0.03 & 0.84 & 0.79 & 0.80 & -3.461 \\ 
\cellcolor{gray!25}MF-X-RK & 0.66 & 0.60 & 0.18 & 0.98 & 0.96 & 0.83 & -3.405 \\ 
\addlinespace
\hdashline
\addlinespace
EW-Comb & 0.27 & 0.07 & 0.01 & 0.56 & 0.57 & 0.34 & -3.429 \\ 
Median-Comb & 0.13 & 0.06 & 0.01 & 0.50 & 0.46 & 0.25 & -3.382 \\ 
\addlinespace
\hdashline
\addlinespace
RS-Comb & 0.50 & 0.11 & 0.04 & 0.71 & 0.81 & 0.93 & -3.472 \\ 
MS-Comb & 0.08 & 0.05 & 0.01 & 0.55 & 0.57 & 0.34 & -3.429 \\ 
\addlinespace
\hdashline
\addlinespace
\cellcolor{gray!25}MCS-Comb & 0.81 & 0.28 & 0.14 & 0.80 & 0.89 & 0.82 & -3.458 \\ 
\cellcolor{gray!25}WL-MCS-Comb & 0.83 & 0.21 & 0.09 & 0.83 & 0.82 & 0.94 & -3.469 \\ 
MCS-RS-Comb & 0.50 & 0.11 & 0.04 & 0.74 & 0.76 & 0.95 & \cellcolor{gray!75}-3.478 \\ 
\cellcolor{gray!25}WL-MCS-RS-Comb & 0.66 & 0.16 & 0.07 & 0.68 & 0.83 & 0.97 & \cellcolor{gray!75}-3.480 \\ 
\cellcolor{gray!25}MCS-MS-Comb & 0.99 & 0.25 & 0.11 & 0.86 & 0.91 & 0.81 & -3.452 \\ 
\cellcolor{gray!25}WL-MCS-MS-Comb & 0.83 & 0.21 & 0.09 & 0.81 & 0.82 & 0.98 & \cellcolor{gray!75}-3.481 \\ 
		\bottomrule
				\end{tabular}
				\begin{tablenotes}[flushleft]
					\setlength\labelsep{0pt}
\item  \textbf{Notes}: Columns from UC to BD-3 report the p-values of the six backtesting procedures described in Table \ref{tab:backtests}. Column $FZLoss$ represents the averages of the $FZLoss$. Light shades of gray denote the success in the backtesting procedures, at significance level $\alpha = 0.05$. Dark shades of gray denote the inclusion in the MCS, at significance level $\alpha = 0.25$. Sample period: 2018-12-03 to 2022-06-28 (797 observations). VaR and ES are calculated at the level $\tau=0.025$. 
				\end{tablenotes}
			\end{threeparttable}
		}
	\end{table} 
	\begin{table}[htbp]
	\centering
	\caption{Summary of successful backtests and MCS inclusions across indices and coverage levels \label{tab:overall_oos_evaluation}}
		\resizebox{0.8\textwidth}{!}{%
			\begin{threeparttable}
				\begin{tabular}{lcccccc}
					\toprule
	
				&	\multicolumn{3}{c}{$\tau=0.025$}& \multicolumn{3}{c}{$\tau=0.01$}\\
							\midrule
						& \#BT & \#MCS 	&  \#BT and \#MCS & \#BT & \#MCS & \#BT and \#MCS\\
						\cline{2-7}
						 \addlinespace 
RM-N & 1 & 2 & 0 & 0 & 2 & 0 \\ 
RM-N-CF & 4 & 5 & 3 & 3 & 5 & 2 \\ 
RM-t & 2 & 3 & 1 & 2 & 5 & 2 \\ 
GARCH-N & 0 & 3 & 0 & 0 & 3 & 0 \\ 
GARCH-N-CF & 5 & 5 & 3 & \cellcolor{green!25}{7} & 6 & 4 \\ 
GARCH-t & 2 & 3 & 1 & 3 & 6 & 1 \\ 
GJR-N & 1 & 5 & 1 & 0 & 3 & 0 \\ 
GJR-N-CF & \cellcolor{green!75}{7} & 6 & 5 & 6 & 5 & 3 \\ 
GJR-t & 2 & 6 & 1 & 4 & 6 & 2 \\ 
RGARCH-RVOL5-N & 0 & 1 & 0 & 0 & 0 & 0 \\ 
RGARCH-RVOL5-N-CF & 2 & \cellcolor{green!25}{8} & 2 & 2 & \cellcolor{green!25}{7} & 2 \\ 
RGARCH-RVOL5-t & 1 & 2 & 1 & 0 & 2 & 0 \\ 
RGARCH-RB-SS-N & 0 & 1 & 0 & 0 & 0 & 0 \\ 
RGARCH-RB-SS-N-CF & 2 & 6 & 2 & 1 & 6 & 1 \\ 
RGARCH-RB-SS-t & 0 & 2 & 0 & 0 & 2 & 0 \\ 
RGARCH-RK-N & 0 & 3 & 0 & 0 & 1 & 0 \\ 
RGARCH-RK-N-CF & 2 & 7 & 2 & 1 & 5 & 1 \\ 
RGARCH-RK-t & 1 & 3 & 1 & 0 & 4 & 0 \\ 
\addlinespace
\hdashline
\addlinespace 
HS-25 & 0 & 0 & 0 & 0 & 0 & 0 \\ 
HS-50 & 0 & 0 & 0 & 0 & 1 & 0 \\ 
HS-100 & 0 & 2 & 0 & 0 & 2 & 0 \\ 
HS-250 & 0 & 3 & 0 & 0 & 3 & 0 \\ 
HS-500 & 0 & 3 & 0 & 1 & 2 & 0 \\ 
\addlinespace
\hdashline
\addlinespace 
SAV & \cellcolor{green!75}{7} & 4 & 3 & 4 & 5 & 2 \\ 
AS & 2 & 4 & 2 & 1 & 3 & 1 \\ 
IG & 5 & 4 & 1 & 4 & 6 & 2 \\ 
CAViaR-X-RVOL5 & \cellcolor{green!25}{6} & 7 & 4 & 5 & \cellcolor{green!75}{8} & \cellcolor{green!25}{5} \\ 
CAViaR-X-RB-SS & \cellcolor{green!25}{6} & \cellcolor{green!25}{8} & 5 & 4 & \cellcolor{green!75}{8} & 4 \\ 
CAViaR-X-RK & \cellcolor{green!25}{6} & 7 & 5 & 5 & \cellcolor{green!25}{7} & \cellcolor{green!25}{5} \\ 
MF-X-RVOL5 & 2 & 5 & 1 & 1 & 3 & 0 \\ 
MF-X-RB-SS & 1 & 5 & 1 & 0 & 3 & 0 \\ 
MF-X-RK & 2 & 3 & 1 & 1 & 2 & 1 \\ 
\addlinespace
\hdashline
\addlinespace 
EW-Comb & 5 & 6 & 3 & 5 & 6 & 3 \\ 
Median-Comb & 4 & 5 & 3 & 4 & \cellcolor{green!25}{7} & 3 \\ 
\addlinespace
\hdashline
\addlinespace 
RS-Comb & 5 & \cellcolor{green!25}{8} & 5 & \cellcolor{green!75}{7} & \cellcolor{green!25}{7} & \cellcolor{green!25}{5} \\ 
MS-Comb & 5 & 6 & 3 & 4 & \cellcolor{green!25}{7} & 4 \\ 
\addlinespace
\hdashline
\addlinespace 
MCS-Comb & \cellcolor{green!25}{6} & \cellcolor{green!25}{8} & 5 & 4 & \cellcolor{green!25}{7} & 3 \\ 
WL-MCS-Comb & 5 & \cellcolor{green!25}{8} & 4 & 4 & 6 & 3 \\ 
MCS-RS-Comb & 5 & \cellcolor{green!75}{\textbf{9}} & 5 & \cellcolor{green!75}{7} & 6 & \cellcolor{green!25}{5} \\ 
WL-MCS-RS-Comb & \cellcolor{green!75}{7} & \cellcolor{green!75}{\textbf{9}} & \cellcolor{green!75}{7} & \cellcolor{green!75}{7} & \cellcolor{green!25}{7} & \cellcolor{green!75}{6} \\ 
MCS-MS-Comb & \cellcolor{green!25}{6} & 7 & 4 & 4 & \cellcolor{green!25}{7} & 3 \\ 
WL-MCS-MS-Comb & \cellcolor{green!25}{6} & \cellcolor{green!75}{\textbf{9}} & \cellcolor{green!25}{6} & 5 & \cellcolor{green!25}{7} & \cellcolor{green!25}{5} \\

		\bottomrule
				\end{tabular}
				\begin{tablenotes}[flushleft]
					\setlength\labelsep{0pt}
\item  \textbf{Notes}:  \#BT denotes the number of indices (out of nine) for which the model passes all six backtests at the 5\% level. \#MCS denotes the number of indices for which the model belongs to the SSM identified by the MCS at the 25\% level. \#BT and \#MCS counts the number of indices for which both criteria hold jointly. Larger values indicate better performance. Dark green shading denotes the best performance in each column, while light green shading indicates the second-best. The maximum attainable value (9) is reported in bold.

				\end{tablenotes}
			\end{threeparttable}
		}
	\end{table} 

\section{Conclusions \label{sec:conclusions}}
\noindent

The continuous evolution of financial markets, together with the growing influence of economic policy actions and geopolitical events, has led to the development of more advanced risk management techniques.
In this framework, VaR and ES remain key measures for both regulatory and risk management purposes.
To mitigate the influence of multiple sources of uncertainty that may affect model performance, this study investigated the effectiveness of forecast combination strategies in improving the predictive accuracy of VaR and ES forecasts generated by individual models.
Our results confirm that, under dynamic market conditions, no single model consistently outperforms the others. To address this issue, we proposed new combination methods based on the MCS methodology, employing a joint VaR-ES loss function to identify the SSM, whose forecasts are then combined using alternative weighting schemes.
An empirical analysis conducted on nine stock market indices at the 2.5\% and 1\% risk levels compared the proposed approaches with 32 individual models, the benchmark mean and median combinations and the RS-Comb and MS-Comb strategies introduced by \cite{Taylor:2020}.
The findings show that the proposed combined predictors outperform the competing methods, achieving satisfactory backtesting results and consistently entering the SSM for both risk levels. In particular, one of the proposed predictors, namely WL-MCS-RS-Comb, enters the SSM for all nine indices considered at the 2.5\% risk level, whereas the standard benchmarks, such as EW-Comb and Median-Comb, enter only six and five times, respectively. Furthermore, WL-MCS-RS-Comb jointly passes all the backtests used in this study and enters the SSM in seven out of nine indices at the 2.5\% risk level, achieving the highest value among all competing specifications.

Future research could expand the range of models considered, for example, by including hybrid non-parametric approaches such as the HS–RiskMetrics specification \citep{Boudoukh:Richardson:Whitelaw:1998}, or quantile regression forest approaches such as that proposed by \citet{Candila:Petrella:2025}, and explore data-driven methods to select the optimal $\lambda$ in the weighted $FZLoss$.

\section*{Acknowledgements}

Alessandra Amendola acknowledges funding from COST Action HiTEc (CA21163), under the framework of COST (European Cooperation in Science and Technology). Giuseppe Storti and Antonio Naimoli acknowledge financial support under the National Recovery and Resilience Plan (NRRP), Mission 4, Component 2, Investment 1.1, Call for tender No. 104 published on 2.2.2022 by the Italian Ministry of University and Research (MUR), funded by the European Union – NextGenerationEU – Project Title: Methodological and computational issues in large-scale time series models for economics and finance (ID: 20223725WE) – CUP  D53D 2300610 0006 - Grant Assignment Decree No. n. 967 of 30/06/2023.
%The authors are also grateful to the participants at SIS 2023 – Statistical Learning, Sustainability and Impact Evaluation – and at the 2nd Italian Conference on Economic Statistics for their valuable comments and feedback. The authors sincerely thank Giovanni De Luca and Giampiero M. Gallo for their insightful discussions and constructive remarks.

\bibliographystyle{chicago}
\bibliography{BIBLIO}

\end{document}